\documentclass{LMCS}
\usepackage{oldlfont,amssymb,latexsym,euscript,stmaryrd}
\usepackage{graphicx}
\usepackage{enumerate}
\usepackage[all]{xy}
\usepackage{hyperref}

\def\[{\ensuremath{[ \! [}}
\def\]{\ensuremath{] \! ]}}
\def\({\ensuremath{( \! [}}
\def\){\ensuremath{] \! )}}

\def\C{{\cal{C}}}

\def\Y{{\mathbf{Y}}}

\def\0{{\mathbf{0}}}

\def\1m{\iota}
\def\restrict{ \upharpoonright}

\def\1{{{\mathbf{1}}}}

\def\restrict{\!\! \upharpoonright\!\!}

\newcommand{\sqleq}{\leq^E}
\newcommand{\seq}{{\mathrm{seq}}}
\newcommand{\zap}{\lightning}

\newcommand{\slambda}{{\mathsf \lambda}}
\newcommand{\strat}{{\mathrm{strat}}}

\newcommand{\ext}{\leq^E}

\newcommand{\str}{{\mathbf{strat}}}

\newcommand{\lomt}{\Lambda^\top_\bot}

\newcommand{\el}{{\mathsf{else}}}

\newcommand{\lra}{\longrightarrow}
\newcommand{\bicp}{{\mathcal{BBC}}}

\newcommand{\pj}{{\mathsf{proj}}}
\newcommand{\cse}{{\mathtt{case}}}

\newcommand{\bist}{{\mathcal{BBO}}}

\newcommand{\proj}{{\mathsf{proj}}}

\newcommand{\ski}{{\mathtt skip}}

\newcommand{\e}{\top}

\newcommand{\spc}{\hspace{2pt}}
\newcommand{\com}{{\Sigma}}

\newcommand{\up}{\updownarrow}
\newcommand{\updnarrow}{\uparrow\!\!\!\downarrow}

\newcommand{\callcc}{{\mathtt{callcc}}}

\newcommand{\nat}{{\mathtt{nat}}}
\newcommand{\id}{{\mathsf{id}}}

\newcommand{\tr}{{\mathrm{tr}}}

\newcommand{\gd}{\Sigma}
\newcommand{\lamw}{\lomt(\omega)}

\newcommand{\inn}{{\mathsf{in}}}

\newcommand{\IF}{{\mathtt{IF0}}}

\newcommand{\If}{{\mathsf{IF0}}}

\newcommand{\inl}{{\mathtt{in_{l}}}}
\newcommand{\inr}{{\mathtt{in_{r}}}}

\newcommand{\Na}{{\mathbb{N}}}

\newcommand{\catch}{{\mathtt{catch}}}

\newcommand{\Cu}{{\EuScript C}}

\newcommand{\pred}{{\mathtt{pred}}}
\newcommand{\pre}{{\mathsf{pred}}}
\newcommand{\then}{{{\mathsf{then}}}}

\newcommand{\suc}{{\mathtt{succ}}}
\newcommand{\scc}{{\mathsf{succ}}}
\newcommand{\fst}{{\mathsf{fst}}}
\newcommand{\snd}{{\mathsf{snd}}}

\newcommand{\inj}{{\mathsf{inj}}}

\newcommand{\iJ}{{\mathsf{inj}}}

\newcommand{\ter}{{\mathbf{1}}}

\def\doi{3 (2:5) 2007}
\lmcsheading%
{\doi}
{1--22}
{}
{}
{Dec.~20, 2005}
{May\phantom{.}~15, 2007}
{}   

\begin{document}
\title[Bistable Biorders]{Bistable Biorders: a Sequential Domain Theory}
\author[J.~Laird]{James~Laird} 
\address{Dept. of Informatics, University of Sussex, UK}
\email{jiml@sussex.ac.uk}
\thanks{Research Supported by EPSRC grant S72191.}
%\date{}
\begin{abstract}We give a simple order-theoretic construction of a Cartesian closed category of sequential functions. It is based on bistable biorders, which are sets with a partial order --- the extensional order --- and a bistable  coherence,  which captures equivalence of program behaviour, up to permutation of top (error) and bottom (divergence). We show that monotone and bistable functions (which are required to preserve bistably bounded meets and joins) are strongly sequential, and use this fact to prove  universality results for the bistable biorder semantics of the simply-typed lambda-calculus (with atomic constants), and an extension with arithmetic and recursion.

We also  construct a bistable model of SPCF, a higher-order functional programming language with non-local control.  We use our universality result for the lambda-calculus to show that the semantics of SPCF is fully abstract. We then establish a direct correspondence between bistable functions and sequential algorithms by showing that sequential data structures give rise to bistable biorders, and that each bistable function between such biorders is computed by a sequential algorithm. 
%We discuss variants on this semantics, such as call-by-value and CPS interpretations, and argue that bistable biorders provide a flexible semantic setting in which intensional properties of programs can be represented faithfully. 
\end{abstract}
\maketitle
\section{Introduction}
Since its inception, domain theory has been a dominant paradigm in denotational semantics; it is a natural and mathematically rich theory with broad applicability across a wide range of phenomena. However, a limitation of domain theory has been its failure to capture the intensional aspects of computation. The observation of Plotkin \cite{Plo}, that the continuous functional model of PCF is not fully abstract, because it  contains functions which are not \emph{sequential}, is symptomatic, but the problem cuts deeper; in the presence of computational effects such as state or concurrency, intensional properties such as the order of computation become critical, and must be captured by some means in any sound model.

Thus, a longstanding problem in domain theory, and the subject of a significant amount of research \cite{be,KP,BC,ehB}, has been to find a simple characterization of higher-order sequential functions which is wholly extensional in character. Typically, what is sought is some form of mathematical  structure, such that all set-theoretic functions which preserve this structure are sequential \emph{and} can be used to construct a Cartesian closed category; the basis for a ``sequential domain theory''. 

Clearly, any solution to this problem is dependent on what one means by sequential. It has been closely associated with the full abstraction problem for PCF, although it is now known that  PCF sequentiality cannot be characterized effectively in this sense \cite{KP,Lo}. %The difficulty of the PCF problem, has led to a \emph{intensional models} of PCF   

Another notion of sequentiality --- the observably sequential functionals --- was discovered by Cartwright and Felleisen \cite{CF}. They observed that if one or more errors  are added to a functional language, then the order of evaluation of programs becomes observable by varying their inputs. Thus each function corresponds to a unique evaluation tree or sequential algorithm \cite{CCF}, which can be reconstructed from its graph. The observably sequential functionals do form a cartesian closed category, which contains a fully abstract model of SPCF --- PCF with errors and a simple control operator. However, the definitions of observably sequential functions and sequential algorithms are based implicitly or explicitly on intensional notions of sequentiality, and hence they cannot offer  a characterization of it in the above sense. So we may refine our original problem to ask whether there is a simple, order-theoretic characterization of observable sequentiality. %, as a

This paper suggests such a characterization.  We will construct a cartesian closed category of biordered sets and order-preserving ``bistable'' functions. We prove that bistable functions correspond to the observably sequential functions both indirectly --- by showing that they may used to give models of observably sequential languages which are \emph{universal} (every element is the denotation of a term) and fully abstract --- and directly, by showing that each sequential data structure yields a bistable biorder, and that every bistable and continuous function between such orders is ``realized'' by a sequential algorithm.

Bistable biorders are  analogous to Berry's bidomains \cite{be,bet}, which combine the extensional order with the stable order. Although the bidomain model of PCF is not sequential, even at first order types, the bidomain model of \emph{unary} PCF (which contains a $\top$ element at each type) is sequential, and universal \cite{ufpc,fi}.  The connection with observably sequential functions is made by viewing top as an error element. Under this interpretation, the monotone and stable functions on bidomains are not observably sequential, because they are not ``error-propagating'' (i.e. sequential with respect to $\top$ as well as $\bot$). However, the duality between $\bot$ and $\top$ suggests that we ``symmetrize'' the stable order, to obtain a notion of \emph{bistable} order.

Bistable coherence may be thought of as ``behavioural equivalence up to the point of failure --- i.e.  % (the denotations of) two programs $M$ and $N$  are in the stable order if they are in the extensional (observational) order, and $N$ performs (at least) all of the computation-steps that $M$ does. By analogy, 
 we may say that $M$ and $N$ are in the bistable order if they are in the extensional order, and $M$ and $N$ perform \emph{the same} computation-steps. 
 $M$ and $N$ are coherent if they behave in a way  \emph{except} that $M$ may diverge where $N$ raises an error, or vice-versa 

Bistable functions are required to preserve the bistable order, and bistably bounded meets and joins. The proof that bistable functions are sequential is surprisingly simple.  Informally, if we have a function which may evaluate two of its components in parallel, we  may consider two arguments  which are identical except that one diverges in the first component, and produces $\top$ in the second, and the other produces $\top$ in the first argument and diverges in the second. These arguments are bounded in the bistable order: their meet diverges in both components. Our function will produce an error when applied to either argument, but will diverge when applied to their meet, and hence it cannot be bistable.

\subsection{Related Work}
The notion of bistable biorder which is elaborated here was first presented (in a slightly different form) in  \cite{cslb}, together with a (different) proof of full abstraction for a model of SPCF. Curien  \cite{cubi}, Streicher \cite{strd} and L\"ow \cite{TL} have  studied bistable functionals, and proved versions of some of the results described here (such as the correspondence between sequential algorithms and bistable functions in \cite{cubi,TL}). The use of definable retractions to prove definability and full abstraction for observably sequential languages originates with Longley \cite{Long,LongU}. The concluding section of this paper gives references to more recent work on bidomain models of sequential languages.
\subsection{Outline of the Paper}
In Section 2, we describe the notion of bistable biorder and bistable function, and prove that it yields a Cartesian closed category. We prove that this contains a universal model of the simply-typed $\lambda$-calculus $\lomt$ over a single atomic type containing two constants ($\top$ and $\bot$), equivalent to the  ``minimal model'' of $\lomt$ \cite{pad}. In Section 3, we develop a notion of complete bistable biorder, or bistable bicpo, and show that  we may define a CCC of bicpos and continuous and bistable functions.  We give a semantics of SPCF in this category, and prove that it is fully abstract. In section 4 we describe a universal model of a $\lambda$-calculus extending $\lomt$ with arithmetic operations and recursion, which may be viewed as a target language for CPS interpretation of observably sequential languages such as SPCF. %We give an instance of such a CPS interpretation of call-by-name SPCF, and show that it is equivalent to a direct interpretation.
In Section 5, we investigate the correspondence between sequential algorithms on sequential data structures and bistable functions, showing that each of the latter gives rise to a bistable bicpo, and that each sequential algorithm on the ``function-space''  computes a bistable function. We then prove that every bistable function is computed in this way, and hence that there is a full embedding of the category of sequential data structures and sequential algorithms in the category of bistable bicpos and bistable and continuous functions.

\section{Bistable Biorders}
%We give the following definition of a bistable biorder.
\begin{defi}A bistable biorder is a tuple $(D, \sqleq, \up)$, where $(D,\sqleq)$ is a partial order (the \emph{extensional} order), and $\up$ is an equivalence relation (\emph{bistable coherence}) on $D$ such that each $\up$-equivalence class is a \emph{distributive} lattice with respect to $\sqleq$, and inclusion into $D$ preserves meets and joins.
\end{defi}
Bistable biorders were introduced in \cite{cslb} as biordered sets (hence the name). In particular, we may define a bistable biorder to be a tuple $\langle D, \sqleq, \leq^B\rangle$, where $(D,\sqleq)$ and $(D,\leq^B)$ are partial orders such that: 
\begin{enumerate}[$\bullet$]
\item $a$ and $b$ are  are bounded above in $\leq^B$ if and only if they are bounded below in $\leq^B$.
\item If $a$ and $b$ are bounded above in $\leq^B$ then there are elements $a \wedge b, a\vee b \in D$ which are (respectively) the  greatest lower bound and least upper bound of $a$ and $b$ with respect to both orders.% (hence $\leq$ is included in $\sqleq$ since if $a \leq b$ then $a \wedge b = a$ and hence $a \sqleq b$).
\item If $\{a,b,c\}$ is bounded above in $\leq^B$, then 
$a \vee (b \wedge c) = (a \vee b) \wedge (a \vee c)$ (and so  $a \wedge (b \vee c) = (a \vee b) \wedge (a \vee c)$. 
\end{enumerate}
\begin{prop}The definitions of bistable biorder are equivalent.
\end{prop}
\proof From the bistable order, we may define the bistable coherence relation: $a \up b$ if $a$ and $b$ are bounded above in $(D,\leq^B)$. This is an equivalence relation, since if $f,g \leq^B p$ and $g,h \leq^B q$, then $g \leq^B p,q$ and hence $p,q$ are bounded above and thus $f \up h$. 

From the bistable coherence relation, we may define the bistable order  $x \leq^B y$ if $x \up y$ and $x \sqleq y$.  
\qed 
%\subsection{Bistable Functions}
We shall now construct a Cartesian closed category of bistable biorders and monotone and bistable functions. 
\begin{defi}
A function $f:D \rightarrow E$ is \emph{monotone} if for all $x,y \in |D|$, $x \sqleq y$ implies $f(x) \sqleq f(y)$
and \emph{bistable}  if for each $x$, $f \restrict\spc  [x]_\up$ is a lattice homomorphism into $[f(x)]_\up$\\
--- i.e. for all $x,y \in |D|$ such that $x\updnarrow y$, $f(x) \up f(y)$,  $f(x \wedge y) = f(x)\wedge f(y)$ and  $f(x \vee y) = f(x)\vee f(y)$. 
\end{defi}
We define a category $\bist$ in which objects are   bistable biorders and morphisms are  monotone and bistable functions.
\begin{lem}$\bist$ is bi-Cartesian.
\end{lem}
\proof
The product and co-product operations on bistable orders are defined directly (pointwise):
%\begin{defi}
\begin{enumerate}[$\bullet$]
\item $A \times B = (|A| \times |B|, \sqleq_A \times \sqleq_B, \up_A \times \up_B)$,
\item $A + B = (|A| + |B|, \sqleq_A + \sqleq_B, \up_A + \up_B)$.
\end{enumerate}
The unit for the product is the one-point biorder, $\ter$  and the unit for the co-product is the empty biorder.
\qed 
We will now show that $\bist$ is Cartesian closed by defining an exponential: a  bistable biorder of functions, in which the extensional order is standard, and the bistable order is a symmetric version of the stable order.
\begin{defi}Given bistable biorders $D,E$, we define the function-space $D \Rightarrow E$ to be the set of  monotone and bistable functions from $D$ to $E$, with
\begin{enumerate}[$\bullet$]
\item $f \sqleq g$ if for all $x \in D$,  $f(x) \sqleq g(x)$,
\item $f \up g$  if for all $x \in D$ $f(x) \up g(x)$, and if $x \up y$ (and hence $f(y) \updnarrow g(x)$) then $f(x) \wedge g(y) = f(y)\wedge g(x)$ and $f(x) \vee g(y) = f(y) \vee g(x)$.
%$f \leq_{D \Rightarrow E} g$ if for all $x \in |D|$ $f(x) \leq_E g(x)$, and if $x \leq_D y$ (and hence $f(y) \updnarrow g(x)$) then $f(x) = f(y)\wedge g(x)$ and $g(y) = f(y) \vee g(x)$.
\end{enumerate}
\end{defi}
\begin{lem}$D \Rightarrow E$ is a bistable biorder.
\end{lem}
\proof
If $f \up g$ then $f(a) \updnarrow g(a)$ for all $a$, and so we may define $\sqleq$ meets and joins  $f \wedge g$ and $f \vee g$ pointwise:\\ 
$(f\wedge g)(a) = f(a) \wedge g(a)$ and $(f\vee g)(a) = f(a) \vee g(a)$.\\
We now show that $f \wedge g$ and $f \vee g$ are monotone and bistable functions  --- e.g. if 
$a \updnarrow b$ then $ (f \wedge g)(a \vee b) = (f \wedge g)(a) \vee (f \wedge g)(b) $. Observe that $f(a) \wedge g(b) = g(a) \wedge f(b) \sqleq f(a),f(b),g(a),g(b)$ and so $f(a) \wedge g(b), f(b) \wedge g(a) \sqleq f(a) \wedge f(b), g(a) \wedge g(b)$. Hence:\\
$(f \wedge g)(a \vee b) = f( a \vee b) \wedge g(a \vee b) = (f(a) \vee f(b)) \wedge (g(a) \vee g(b)) = (f(a) \wedge g(a)) \vee (f(a) \vee g(b)) \vee (f(b) \wedge g(a)) \vee (f(b) \wedge g(b)) = (f(a) \wedge g(a)) \vee (f(b) \wedge g(b)) =   (f \wedge g)(a) \vee (f \wedge g)(b)$. 

Next, we show that $f \up f \wedge g$ and $f \up f \vee g$:\\
For all $x$, $f(x) \up (f \wedge g)(x) = f(x) \wedge g(x)$, and for all $y$ such that $x \up y$,
$f(x) \wedge (f \wedge g)(y) = f(x) \wedge f(y) \wedge g(y) = f(y) \wedge f(x) \wedge g(x) =   f(y) \wedge (f \wedge g)(x)$ and \\
$ f(x) \vee (f \wedge g)(y) = f(x) \vee (f(y) \wedge g(y)) = (f(x) \vee f(y)) \wedge (f(x) \vee g(y)) = (f(y) \vee f(x)) \wedge (f(y) \vee g(x)) =  f(y) \vee (f \wedge g)(x)$. 

Finally, we need to prove that $\updnarrow$ is transitive, for example, suppose $f \updnarrow g$ and $g \updnarrow h$. Suppose $x \updnarrow y$. Then:\\ $f(x) \wedge h(y) = f(x) \wedge h(y) \wedge (f(x) \vee g(y))  = f(x) \wedge h(y) \wedge (f(y) \vee g(x)) \\= (f(x) \wedge h(y) \wedge f(y)) \vee (f(x) \wedge h(y) \wedge g(x)) \\\sqleq f(y) \vee (f(x) \wedge g(y) \wedge h(x)) = f(y) \vee (f(y) \wedge g(x) \wedge h(x)) = f(y)$.
\\ Similarly, $f(x) \wedge h(y) \sqsubseteq   h(x)$, $f(y) \wedge h(x) \sqsubseteq   f(y)$ and   $f(y) \wedge h(x) \sqsubseteq   h(x)$. So $f(x) \wedge h(y) = f(y) \wedge h(x)$. By duality, $f(x) \vee h(y) = f(y) \vee h(x)$ and so $f \updnarrow h$ as required.
\qed 
\begin{prop}$(\bist,\ter,\times)$ is cartesian closed.
\end{prop}
\proof
 We need to show that the natural bijection  taking $f:A \times B \rightarrow C$ to $\Lambda(f):A \rightarrow (B \Rightarrow C)$ such that $\Lambda(f)(a)(b) = f(\langle a,b \rangle)$, and its inverse,  are well-defined on bistable biorders and bistable functions. This is similar to the proof for (stable) biorders  and stable and monotone functions \cite{bet}.

For example, to show that $\Lambda(f)$ preserves bistable coherence:\\
Suppose $a \up _A a'$.
Then for all $b \up_B b'$, $\langle a,b \rangle \up \langle a',b'\rangle$, and e.g.  $\Lambda(f)(a)(b) \wedge \Lambda(f)(a')(b') =  f(\langle a,b \rangle \wedge \langle a',b'\rangle) = f(\langle a,b' \rangle) \wedge f(\langle a',b\rangle) = \Lambda(f)(a)(b')\wedge \Lambda(f)(a')(b)$.  Similarly, $\Lambda(f)(a)(b) \vee \Lambda(f)(a')(b') = \Lambda(f)(a)(b')\vee \Lambda(f)(a')(b)$ and hence $\Lambda(f)(a) \up \Lambda(f)(b')$ as required. 

Conversely, to show that if $g:A \rightarrow (B \Rightarrow C)$ is bistable, then  $\Lambda^{-1}(g)$ is bistable, suppose $\langle a,b \rangle \up_{A \times B} \langle a',b' \rangle$. Then $a \up a'$ and $b \up b'$ and by bistability of $g$, $g(a)(b) \wedge g(a')(b') =  g(a)(b') \wedge g(a')(b)$ and  $g(a)(b) \vee g(a')(b') =  g(a)(b') \vee g(a')(b)$. So e.g.   $\Lambda^{-1}(g)(\langle a,b \rangle \wedge \langle a',b' \rangle) = g(a \wedge a')(b \wedge b') = g(a)(b) \wedge g(a)(b') \wedge g(a')(b) \wedge  g(a')(b') = g(a)(b) \wedge g(a')(b') = \Lambda^{-1}(g)(\langle a,b \rangle) \wedge \Lambda^{-1}(g)(\langle a',b' \rangle)$.
\qed

A bistable biorder $D$ is \emph{pointed} if $(D,\sqleq)$ has a least element $\bot$ and a greatest element $\top$, such that $\bot \up \top$. A monotone   bistable function $f$ of pointed biorders is \emph{bistrict} if it preserves the meet and join of the empty set --- i.e. $f(\top) = \top$ and $f(\bot) = \bot$. We define the category $\bist_s$ of pointed bistable biorders and strict, monotone and bistable functions.
\begin{prop}The inclusion of $\bist_s$ into $\bist$ has a  left adjoint.\end{prop}
\proof
The \emph{bilifting} operation takes a bistable biorder $A$ to a pointed bistable biorder by adding \emph{two} new points, $\top$ and $\bot$: $A^\top_\bot = (A \times \{*\}) \cup \{\bot,\top\} $, where: 
 \begin{enumerate}[$\bullet$]
\item $x \sqleq y$ if  $x = \bot$ or $y = \top$, or $x = \langle x',*\rangle, y  = \langle y',*\rangle$ and  $x' \sqleq y'$,
\item $x \up y$ if $x,y \in\{\bot,\top\}$  or $x = \langle x',*\rangle, y  = \langle y',*\rangle$ and  $x' \up y'$.
 \end{enumerate} 
For any pointed $B$, $\bist(A,B) \cong \bist_s(A^\top_\bot,B)$.
\qed 
\subsection{First-Order Sequentiality and Universality}
A key step in proving universality for observably sequential languages  is the observation that the monotone and bistable functions on pointed bistable biorders are \emph{bisequential} (i.e. sequential with respect to both $\bot$ and $\top$ elements).% in the sense of Milner-Vuillemin. 
\begin{defi}Given pointed  bistable  biorders $A_1,\ldots, A_n,B$,  a function $f:A_1 \times \ldots \times A_n \rightarrow B$ is $i$-strict if $\pi_{i}(x) = \bot$ implies $f(x) = \bot$   and $\pi_{i}(x) = \top$ implies  $f(x) = \top$.
\end{defi}
\begin{lem}\label{seq}Given pointed  bistable  biorders $A_1,\ldots, A_n$,  every strict, monotone and bistable function  $f:A_1 \times \ldots \times A_n \rightarrow \Sigma$ is $i$-strict for some $i \leq n$.
\end{lem}
\proof%If $B$ is the one-point order then $f$ is the terminal map, which is $i$-strict for all $i$. So we assume henceforth that $B$ has at least two elements.

Given $j \leq n$, let $\bot[\top]_j = \langle x_i\ |\ i \leq n\rangle$, where $x_i = \top$ if $i =j$, and $x_i = \bot$ otherwise.  Similarly $\top[\bot]_j = \langle x_i\ |\ i \leq n\rangle$, where $x_i = \bot$ if $i =j$, and $x_i = \top$ otherwise. 

%Let $\bot_j = \langle x_i\ |\ i \in I\rangle$, where $x_i = \bot$ if $i =j$, and $x_i = \top$, otherwise. Let $\top_i = \neg \bot_i$.\\
If $\pi_i(x) = \bot$ then $x \sqleq\top[\bot]_i$, and if $\pi_i(x) = \top$, $\bot[\top]_i \sqleq x$. Thus $f$ is $i$-strict if $f(\top[\bot]_i) = \bot$ and $f(\bot[\top]_i) = \top$.
Since $\bot \leq^B \top$, we have $\top[\bot]_j\updnarrow \top[\bot]_k$ for all $j,k \leq n$, and $\bigwedge_{i \leq n}\top[\bot]_i  = \bot$. Hence $\bigwedge_{i \leq n}f(\top[\bot]_j) =  f(\bigwedge_{i \leq n}\top[\bot]_i) = f(\bot) = \bot$, and so for some $i$, $f(\top[\bot]_i)   = \bot$.  Similarly $f(\bigvee_{i \in I}(\bot[\top]_i)) = \top$, and so $f(\bot[\top]_j) = \top$ for some $j$. Moreover, if $i \not = j$, then $\bot[\top]_j \sqleq \top[\bot]_i$, and so either $\bot[\top]_j = \top[\bot]_i$ --- in which case each $A_k$ is the one-point order --- or else $i = j$ as required,  and hence $i$ is unique --- i.e. bisequential functions are \emph{strongly sequential}.
\qed 

\subsection{Universality for $\lomt$}
Let $\lomt$  be the simply-typed $\lambda$-calculus with products, over a single base type $\Sigma$ containing the constants $\top$ and $\bot$. 
%standard axioms for the product --- $\pi_i\spc \langle x_1,x_2 \rangle = x_i$ and $\langle \pi_1\spc x,\pi_2\spc x \rangle = x$ --- and
 The ``minimal''  model  of this language (that is, the model inducing the maximal consistent theory containing $\beta$ and $\eta$)  was shown to be effectively presentable by Padovani \cite{pad} using an analysis of the syntax.
By Cartesian closure of $\bist$, we obtain a model of $\lomt$ in which each type is interpreted as the corresponding bistable biorder. %\footnote{We will also use the $\lambda$-calculus for describing elements of this model}. 
We will show that this is the minimal model. 
%By fixing an interpretation of the ground types, we obtain an interpretation of the simply typed $\lambda$-calculus in $\bist$.  % with function-types $\sigma \Rightarrow T$ and product types $\sigma \times T$. Thus

For each type $S$ of $\lomt$, an element of the corresponding biorder is \emph{definable} if it is the denotation of a closed term of type $S$.   Universality holds at $S$ if every  element of $S$  
is definable. Universality at \emph{first-order} function types  is a consequence of sequentiality.
\begin{lem}\label{fst}The bistable model of $\lomt$ is universal at all types of the form   $\Sigma^n \Rightarrow \Sigma^m$. 
\end{lem}
\proof Suppose $m= 1$. If $f$ is constant ($\top$ or $\bot$), then $f$ is definable. Otherwise, $f$ is strict, and hence for some $i$, $f$ is $i$-strict --- i.e. $f  = \pi_i$, and is therefore definable.  
If $m >1$, we have $f = \langle f;\pi_i \ |\ i \leq m \rangle$, and $f;\pi_i$ is definable for each $i$ and so $f$ is definable. 

\qed 
We will now prove that universality at higher-order types \emph{reduces} to universality at first-order, using the notion of \emph{definable retraction}. 
\begin{defi}Given types $S,T$, a definable retraction from $S$ to $T$ (which we may write $\inj:S \unlhd T:\proj$ or just $S \unlhd T$) is a pair of  terms: $\inj:S \Rightarrow T$ and $\proj:T \Rightarrow S$ which denote a retraction  in $\bist$ (i.e. $\[\inj\];\[\proj\] = \id_S$).
\end{defi}
% By the following lemma. 
\begin{lem}\label{ret}If  universality holds at type $T$, and  $\inj:S \unlhd T:\proj$, then universality holds at type $S$.
\end{lem}
\proof 
Given an element $e \in S$, we have a term $M:T$ such that $\[M\] = e;\iJ$ and thus $\[\proj\spc M\] = e;\iJ;\pj = e$.  
\qed 
So we can prove universality for $\lomt$ by showing that every $\lomt$ type is a definable retract of a first order type. To do so, we require a few simple facts about definable retractions.%; for example, $T$ is a definable retract of $\sigma \times T$, and $\sigma \Rightarrow T$.% and the following Lemma.
\begin{lem}\label{retrac1}If $\inj_T:T_1 \unlhd T_2:\proj_T$,  and $\inj_S:S_1 \unlhd S_2:\proj_S$, then 
$S_1 \Rightarrow T_1 \unlhd S_2 \Rightarrow T_2$ and $S_1 \times T_1  \unlhd
S_2 \times T_2$.
\end{lem}
\proof We have, for example, $\lambda fx.\inj_T\spc (f\spc (\proj_S\spc x): S_1 \Rightarrow T_1 \unlhd S_2 \Rightarrow T_2:\lambda fx.\proj_T\spc (f\spc(\inj_S x))$.
\qed 
The key to reducing the order of the function-space is the fact that for any $n$, $(\Sigma^n \Rightarrow \Sigma) \Rightarrow \Sigma$ is a definable retract of $(\Sigma \Rightarrow \Sigma) \times \Sigma^n$. % as we shall now show.% will now show that for any $n$
\begin{lem}\label{lL1}If  $f:(\com^n \Rightarrow \com) \rightarrow \com$ is a strict bistable function, then for all $e \in \com^n \Rightarrow \com$, $f\spc e = e\spc \langle f\spc\pi_i\ |\ 1\leq i \leq n\rangle$.
\end{lem}
\proof If $e = \top$ then $f\spc e = \top$ by strictness of $f$, and $e\spc \langle f(\pi_i)\ |\ 1\leq i \leq n\rangle = \top$. Similarly, if  $e = \bot$ then $f\spc e = e\spc \langle f\spc \pi_i\ |\ 1\leq i \leq n\rangle = \bot$. Otherwise, $e = \pi_i$ for some $1 \leq i \leq n$, and $e\spc \langle f\spc \pi_i\ |\ 1\leq i \leq n\rangle =  f\spc \pi_i = f\spc e$ as required. 
\qed 
\begin{lem}\label{ll2}Let $g = \slambda x.\slambda y.(x\spc \slambda z.(y\spc \langle x\spc \pi_i\ |\ 1 \leq i \leq n \rangle))$\footnote{Here we are using $\lambda$-calculus notation to describe an element of $\bist$.}. Then $g = \id_{(\com^n \Rightarrow \com) \Rightarrow \com}$ 
\end{lem}
\proof 
We show that for any element $f \in (\com^n \Rightarrow \com) \Rightarrow \com$, $g(f) = f$.  We first note that $g (\top) = \top$ and $g(\bot) = \bot$.

If $f \not = \top$ and $f \not = \bot$, then $f$ is strict (since
$f(\bot) \sqleq f(\top)$). Given $e \in \com^n \Rightarrow \com$,
suppose $f (e) = \top$, then $(g\spc f)(e) = f (\slambda z.(e\spc
\langle f\spc \pi_i\ |\ 1 \leq i \leq n \rangle)) = f\spc (\slambda
z.f(e)) = f (\slambda z.\top) = \top = f(e)$, by Lemma \ref{lL1} and
strictness of $f$, and similarly if $f(e) = \bot$, then $(g\spc f)(e)
= \bot$. Hence $g(f) = f$ as required.  \qed

\begin{lem}\label{ret1} For any $n \geq 1$, $(\com^n \Rightarrow \com) \Rightarrow \com$ is a definable retract of $(\com \Rightarrow \com) \times \com^n$ in $\bist$. 
\end{lem}
\proof Consider the terms $$\inj:((\com^n \Rightarrow \com) \Rightarrow \com) \Rightarrow ((\com \Rightarrow \com) \times  \com^i) = \lambda f.\langle \lambda x.(f\spc \lambda y.x),\langle f \spc \pi_i\ |\ 1 \leq i \leq n\rangle \rangle$$
$$\proj: ((\com \Rightarrow \com) \times  \com^n) \Rightarrow (\com^n \Rightarrow \com) \Rightarrow \com = \lambda x.\lambda g.(\pi_1(x)\spc (g\spc \pi_2(x)))$$
We have $\lambda x.\proj\spc (\inj\spc x) =_{\beta\pi} \lambda x.\lambda y.x\spc \lambda z.y\spc \langle x\spc \pi_i\ |\ 1 \leq i \leq n \rangle$, and hence by Lemma \ref{ll2}, $\[\lambda x.\proj\spc (\inj\spc x)\] = \id_{(\com^n \Rightarrow \com) \Rightarrow \com}$.
\qed 
\begin{lem} \label{ret3}For any $n,m \geq 1$, $(\com^n \Rightarrow \com^m) \Rightarrow \com$ is a definable retract of $\com^{n+m} \Rightarrow \com^{(2n)^m}$.\end{lem}
\proof  By induction on $m$. For the  base case ($m =1$), we have $(\com^n \Rightarrow \com) \Rightarrow \com \unlhd (\com \Rightarrow \com)\times \com^n$ by Lemma \ref{ret1}, and  since $\com \Rightarrow \com \unlhd \com^n+1 \Rightarrow \com$, and $\com \unlhd \com^n \Rightarrow \com$, so $\com^n \unlhd (\com \Rightarrow \com)^n$, we have $(\com^n \Rightarrow \com^m) \Rightarrow \com \unlhd  ((\com^{n+1} \Rightarrow \com)^n)^2 \cong ((\com^{n+1} \Rightarrow \com)^{2n^m}$. % \times  (\com^n+1 \Rightarrow \com)^n \\
For the induction case,\\  $(\com^n \Rightarrow \com^{m+1}) \Rightarrow \com \cong (\com^n \Rightarrow \com^m) \Rightarrow ((\com^n \Rightarrow \com) \Rightarrow \com)$\\ $\unlhd (\com^n \Rightarrow \com^m) \Rightarrow ((\com \Rightarrow \com) \times \com^n)$ by Lemma \ref{ret1}.\\
$\cong (\com \Rightarrow (\com^n \Rightarrow \com^m) \Rightarrow \com) \times ((\com^n \Rightarrow \com^m) \Rightarrow \com)^n$\\
$\unlhd (\com \Rightarrow (\com^{n+m} \Rightarrow \com^{(2n)^{m}})) \times (\com^{n+m} \Rightarrow \com^{(2n)^{m}})^n$ by induction hypothesis\\
$\unlhd (\com^{n+m+1} \Rightarrow \com^{(2n)^{m}})^n \times (\com^{n+m + 1} \Rightarrow \com^{(2n)^{m}})^n$\\
$\unlhd  \com^{n+m+1} \Rightarrow \com^{(2n)^{m}\cdot 2n} \cong \com^{n+m+1} \Rightarrow \com^{(2n)^{m+1}}$ as required. 
\qed   
\begin{lem}\label{lem1}For any type $T$ there exists $n(T),m(T) \in \Na$ such that $T$ is a definable retract of $\com^{n(T)} \Rightarrow \com^{m(T)}$. 
\end{lem}
\proof  is by induction on type structure. For the induction cases: 
%For the base case, we have $\com \unlhd \com^1 \Rightarrow \com^1$. \\
$S \times T \unlhd (\Sigma^{n(S)} \Rightarrow \Sigma^{m(S)}) \times (\Sigma^{n(T)} \Rightarrow \Sigma^{m(T)}) \unlhd \com^{\max\{n(S),n(T)\}} \Rightarrow \Sigma^{m(S)+ m(T)}$.\\
$S \Rightarrow T \unlhd (\Sigma^{n(S)} \Rightarrow \Sigma^{m(S)}) \Rightarrow (\Sigma^{n(T)} \Rightarrow \Sigma^{m(T)}) \\ \cong  (\Sigma^{n(T)} \Rightarrow (\Sigma^{n(S)} \Rightarrow \Sigma^{m(S)}) \Rightarrow \Sigma)^{m(T)}\\   \unlhd (\Sigma^{n(T)} \Rightarrow  (\Sigma^{n(S)+m(S)} \Rightarrow \com^{(2n(S))^{m(S)}})^{m(T)}\\ \cong \Sigma^{n(T)+ n(S)+m(S)} \Rightarrow \Sigma^{(2n(S))^{m(S)} \cdot m(T)}$.
\qed 
By applying Lemma \ref{ret} to Lemmas \ref{fst} and \ref{lem1} we have established:
\begin{thm}\label{theo}The bistable model of $\lomt$ is universal at all types.
\end{thm} 
\begin{cor}The bistable model is minimal.
\end{cor}
\proof It is straightfoward to use universality to show that if $\[M\] \not = \[N\]$, then there is an (applicative) context such that $C[M] =_{\beta\eta\pi} \top$ and $C[N] =_{\beta\eta\pi} \bot$, or vice-versa. Hence any compatible theory containing $\beta\eta\pi$ as well as $M = N$ also contains $\bot = \top$.
\qed 
Our proof  also yields a solution to a related problem: to give a simple axiomatization of the theory of the minimal model. 
\begin{defi} Let  the theory $=_\bot^\top$ over the terms of $\lomt$  be the compatible, symmetric and transitive  closure of $\beta\eta\pi$-equivalence  extended with the axioms $f:(\Sigma^n \Rightarrow \Sigma) \Rightarrow \Sigma = \lambda h.f\spc \lambda x.h\spc \langle f\spc \pi_i\ |\ 1 \leq i \leq n\rangle$ for each $n$. 
\end{defi}
For each type $T$, we have a definable retraction $\inj_T: T \unlhd \com^{n(T)} \Rightarrow \com^{m(T)}: \proj_T$.
\begin{lem}$\proj_T \spc (\inj_T\spc x) =_\bot^\top \lambda x.x$.  
\end{lem}
\proof  This is by induction on $T$, following the definition of $\proj_T$ and $\inj_T$, since to prove that they define a retraction in $\bist$, we used only standard properties of all CCCs (i.e. $\beta\eta\pi$-equivalence) together with  $(\Sigma^n \Rightarrow \Sigma) \Rightarrow \Sigma \unlhd (\Sigma \Rightarrow \Sigma) \times \Sigma^n$.% the additional axioms are sufficient to prove that this is a retraction.  
\qed 
\begin{prop}$M:T =_\bot^\top N:T$ if and only if $\[M\] = \[N\]$.
\end{prop}
\proof 
From left-to-right, this follows from the soundness of the theory $=_\bot^\top$ in the bistable model of $\lomt$.

To prove the converse, suppose  $\[M\] = \[N\]$. Then $\[\inj\spc M\] \in \Sigma^{n(T)} \Rightarrow \Sigma^{m(T)} = \[\inj\spc N\] \in \Sigma^{n(T)} \Rightarrow \Sigma^{m(T)}$. Hence for each $j \leq m(T)$, the terms $\lambda x.\pi_j((\inj\spc M)\spc x)$ and $\lambda x.\pi_j((\inj\spc N)\spc x)$ have  the same \emph{head-normal form} (i.e. $\lambda x.\top$, $\lambda x.\bot$ or $\lambda x.\pi_i\spc x$ for some $1 \leq i \leq n(T)$). Thus  $\inj\spc M  =_\bot^\top  \inj\spc N$ and so $M =_\bot^\top \proj\spc (\inj\spc M) =_\bot^\top \proj\spc (\inj\spc N) =_\bot^\top N$ as required.
\qed  
\section{Bistable bicpos}
We shall now extend our notion of bistable biorder with  notions of completeness and continuity.
\begin{defi}Given $\sqleq$-directed sets $X,Y$, we say that  $X \updnarrow Y$ if for all $x \in X$ and $y \in Y$ there exists $x' \in X$ and $y' \in Y$ such that $x\sqleq x'$, $y \sqleq y'$ and $x' \up y'$. A \emph{bistable bicpo} is a bistable biorder $D$ such that $(|D|,\sqleq)$ is a cpo and  if  $X \updnarrow Y$ then  $\bigsqcup X \updnarrow \bigsqcup Y$ and $\bigsqcup X \wedge \bigsqcup Y = \bigsqcup \{x \wedge y\ |\ x \in X\wedge y \in Y\wedge x \up y\}$
\end{defi}
%Hence any bistable function which is continuous with respect to $\sqleq$ must also be continuous with respect to $\leq$. 
Let $\bicp$ be the category of bistable bicpos and  continuous and bistable functions.
\begin{prop}$(\bicp,\ter,\times)$ is  Cartesian closed.
\end{prop}
\proof 
We show that for any directed set $F$ of functions from $A$ to $B$, a bistable and continuous least upper bound can be defined pointwise --- $(\bigsqcup F)(a) = \bigsqcup F(a)$, where $F(a) =  \{f(a) \ |\ f \in F\}$.

$\bigsqcup F$ is bistable: if $a \updnarrow b$, then we have $F(a) \updnarrow F(b)$ and hence $(\bigsqcup F)(a) \updnarrow (\bigsqcup F)(b)$, and $(\bigsqcup F)(a \vee b) = \bigsqcup \{f(a) \vee f(b)\ |\ f \in F\} \sqleq (\bigsqcup F)(a)\vee (\bigsqcup F)(b)$ 

To show preservation of glbs, we note that $\bigsqcup \{f(a)\wedge g(b)\ |\ f,g \in F\wedge f(a) \up g(b)\} = \bigsqcup \{f(a) \wedge f(b)\ |\ f \in F\}$ by directedness of $F$ (for any $f,g$ such that $f(a) \updnarrow g(b)$, we choose $h$ such that $f,g \sqleq h$ and hence $f(a) \wedge g(b) \sqleq h(a) \wedge h(b)$. Thus 
$(\bigsqcup F)(a)\wedge (\bigsqcup F)(b) = \bigsqcup \{f(a)\wedge g(b)\ |\ f,g \in F\wedge f(a) \up g(b)\} = \bigsqcup \{f(a) \wedge f(b)\ |\ f \in F\} = (\bigsqcup F)(a \wedge b) $.

Now given directed sets of bistable functions $F,G$ such that $F \updnarrow  G$
\begin{description}
\item [$\bigsqcup F \updnarrow \bigsqcup G$] For all $x$, $F(x) \updnarrow G(x)$, and hence  $(\bigsqcup F)(x) \updnarrow (\bigsqcup G)(x)$. 
Now suppose $x \updnarrow y$ --- we need to show that $(\bigsqcup F)(x) \wedge (\bigsqcup G)(y) = (\bigsqcup F)(y) \wedge (\bigsqcup G)(x)$. By symmetry it suffices to show $(\bigsqcup F)(x) \wedge (\bigsqcup G)(y) = \bigsqcup\{f(x) \wedge g(y)\ |\ f \in F\wedge g\in G\wedge f(x)\up g(y)\}  \sqleq (\bigsqcup F)(y)$. 
Given  $f \in F$ and  $g\in G$ such that $f(x)\up g(y)$, there exists $f' \in F$ and $g' \in G$ such that $f\sqleq f',g\sqleq g'$ and  $f' \updnarrow g'$. Hence $f(x) \wedge g(y) \sqleq f'(x)\wedge g'(y) \sqleq f(y)$ and so $f(x)\up g(y) \sqleq (\bigsqcup F)(y)$ as required.
\item [$\bigsqcup F \wedge \bigsqcup G = \bigsqcup\{f \wedge g\ |\ f \updnarrow g\}$] For all $x$, $(\bigsqcup F \wedge \bigsqcup G)(x) =  (\bigsqcup F)(x) \wedge (\bigsqcup G)(x) = \bigsqcup \{f(x) \wedge g(x)\ |\ f(x)\up g(x)\} = \bigsqcup\{f(x) \wedge g(x)\ |\ f \updnarrow g\}$, since for any $f,g$ such that $f(x) \up g(x)$ there exists $f'\in F,g'\in G$ such that  $f' \updnarrow g'$  and $f\sqleq f'$, $g\sqleq g'$, and so $f(x)\wedge g(x) \sqleq f'(x)\wedge g'(x)$. 
\end{description}
\qed 
The  bistable bicpos are also closed under the lifting and coproduct operations.
\subsection{SPCF}

We have defined a cpo-enriched Cartesian closed category of sequential functionals, in which we may interpret PCF. We will now show that we have a fully abstract semantics of  SPCF \cite{CF} --- PCF with a non-local control operator --- $\catch$ --- and  an ``error'', $\top$.  Thus we may connect our bistable semantics of $\lomt$ to the ``original'' observably sequential language, SPCF. In doing so, we establish indirectly the correspondence between observably sequential functionals and bistable functionals, since both yield fully abstract imodels of SPCF.  In the case of the bistable model, our proof of universality for $\lomt$ gives an easy proof of full abstraction, since every SPCF type-object is a limit for a chain of $\lomt$ types. 

%We may then prove of full abstraction is based on the fact that we may view $\lomt$ as a fragment of SPCF, by including $\Sigma$ as a ground type (representing an ``empty'' type, or return type for continuations). This  simplifies the operational semantics, and universality for $\lomt$ yields a very simple proof of full abstraction for SPCF, since every SPCF type-object is a limit for a chain of $\lomt$ types. 

The types of SPCF are given by the following grammar:
$$S,T::= \Sigma\ |\ \nat\ |\ S \Rightarrow T\ |\ S \times T$$
Terms are obtained by extending the simply-typed $\lambda$-calculus with pairing and projection and the following constants:
\begin{description}
\item [Divergence and Error] $\top,\bot:\Sigma$,
\item [Numerals] ${\mathtt{0}}:\nat$, $\suc, \pred:\nat \Rightarrow \nat$, 
\item [Conditionals]$\IF:\nat \Rightarrow (T \times T) \Rightarrow T$, where $T \in \{\Sigma,\nat\}$,
\item [Fixpoints] $\Y:(T \Rightarrow T) \Rightarrow T$
\item [Control] $\catch_n:(\Sigma^n \Rightarrow \Sigma)\Rightarrow \nat$ 
\end{description}
The control operator $\catch$ is a basic form of Cartwright and Felleisen's $\catch$ \cite{CF}; it sends $i$-strict functions ($i$th-projection) to ${\mathtt i}$. Despite its simplicity, it can be used to derive (call-by-name versions of) control operators such as Felleisen's idealized call-with-current-continuation operator $\C:((\nat \Rightarrow \Sigma) \Rightarrow \Sigma) \Rightarrow \nat$ \cite{FD}:
$$\C \equiv \Y \lambda f.\lambda g.((\IF\spc (\catch_2\spc \lambda x.g \spc \lambda y.(\IF\spc y)\spc x))\spc \langle 0,\suc\spc (f\spc \lambda h.g\spc (\lambda z.h \spc (\pred\spc z))) \rangle)$$
(So $\catch_2$ is sufficient to express $\catch_n$ for any $n$.)

We may give a simple operational semantics for SPCF programs --- \emph{closed terms of type $\Sigma$} --- using \emph{evaluation contexts},  
\begin{defi}Evaluation contexts of SPCF are given by the following grammar: 
$$E[\cdot]::= [\cdot]\  |\ E[\cdot]\spc M\ |\ \IF\spc E[\cdot]\ |\  \pi_{i}\spc E[\cdot]\ |\ \suc\spc E[\cdot]\ |\ \pred\spc E[\cdot]$$
\end{defi}
The ``small-step'' operational semantics of SPCF programs is given in Table 1. The rule for $\catch$ makes its connection with control operators such as $\callcc$ apparent; the current continuation (represented as a tuple of evaluation contexts filled with the possible values for $\catch \spc M$) is passed as an argument to $M$. 
\begin{table}
\begin{center}
$E[\top] \lra \top$\\
$E[(\lambda x.M)\spc N] \lra E[M[N/x]]$\\
$E[\pi_i\spc \langle M_1,M_2\rangle] \lra E[M_i]$\\
$E[\pred\spc (\suc\spc {\mathtt n})] \lra E[{\mathtt n}]$\\
$E[\IF\spc {\mathtt{0}}] \lra E[\pi_1]$\\
$E[\IF\spc ({\mathtt \suc\spc n})] \lra E[\pi_2]$\\
$E[\catch_n\spc M] \lra M\spc \langle E[{\mathtt 0}], \ldots E[{\mathtt{n-1}}]\rangle$\\
$E[\Y\spc M] \lra E[M\spc (\Y M)]$\\
\end{center}
\caption{``Small-step'' operational semantics for SPCF programs.}
\end{table} 
For a program $M$ we write $M \Downarrow$ if  $M \twoheadrightarrow \top$. We adopt a standard definition of observational approximation and equivalence:
 given terms $M,N:T$, $M \lesssim N$ if for all compatible program contexts $C[\cdot]$, $C[M]\Downarrow$ implies $C[N]\Downarrow$.

\subsection{The bistable model of SPCF}
The ground type $\nat$ is interpreted as  $\Na^\top_\bot$, where $\Na$ is the set of natural numbers with the trivial extensional and bistable orderings. % ``biflat'' domain with $\bot$ and $\top$ elements, and distinct elements $\underline{i}$ for each $i \in \Na$. 
We interpret $\catch_n$ as the strict bistable function from $\Sigma^n \Rightarrow \Sigma$ to $\Na^\top_\bot$  which sends the $i$th projection to the  value  $i$. The interpretation of the remainder of the language (i.e. PCF) is standard, since $\bicp$ is a cpo-enriched Cartesian closed category.
\begin{prop}\label{snda}$M\Downarrow$ if and only if $\[M\]  = \top$. 
\end{prop}
\proof To show soundness, we need simply to verify that if $M \lra N$  then $\[M\] = \[N\]$. This is standard for all the rules except those for $\top$ and $\catch$. To establish these cases, we prove by induction that evaluation contexts are interpreted as strict maps  --- i.e.  $\[E[\top]\] = \top$ and $\[E[\bot]\] = \bot$. Thus for any closed term $M:\Sigma^n \Rightarrow \Sigma$, if $\[M\]$ is constant, then $\[E[\catch\spc M]\] = \[M\spc E[{\mathtt 0}],\ldots, E[{\mathtt n-1}]\rangle\]$, whilst if $\[M\] = \pi_i$ then $\[M\spc \langle E[{\mathtt 0}],\ldots,E[{\mathtt n-1}]\rangle\] = \[E[{\mathtt{i}}]\] = \[E[\catch\spc M]\]$ 

Adequacy is proved using a Tait-style computability predicate argument as for PCF \cite{Plo}.
\qed 

We prove full abstraction by reduction to universality for $\lomt$. The key to doing this is the observation that for each $i$, the type $\Sigma^i \Rightarrow \Sigma$ is a definable retract of $\nat$. 
%For each $n$, we have injections from $\Sigma^i \Rightarrow \Sigma$ to $\Na^\top_\bot$ sending the $i$-th projection to the integer $i-1$: these are definable as a family of operators $\catch_i:(\Sigma^i \Rightarrow \Sigma) \Rightarrow \nat$ generalizing $\catch$:  
%$$\catch_{n+1} = \lambda f.\C \lambda g.f \spc \langle g\spc 0,g\spc 1, \ldots , g\spc n \rangle$$  

For each $n \geq 1$ we have projection maps from $\Na^\top_\bot$ to $\Sigma^n \Rightarrow \Sigma$ sending $i<n$ to the $i+1$th projection, and $i \geq n$ to $\bot$. These are definable as $n$-ary case statements $\cse_n$, where $\cse_1 = \lambda x.\lambda y. ((\IF\spc x)\spc \langle x,\bot\rangle)$, and 
$$\cse_{n+1} = \lambda x.\lambda y.(\IF\spc x) \langle \pi_1\spc y,(\cse_n\spc (\pred\spc x))\spc (\pi_2\spc y) \rangle$$
\begin{lem}\label{snb}For each SPCF type $S$ there is a sequence of $\lomt$ types $\{S_i\ :\ i \in \omega\}$ with SPCF-definable retractions: $\inj_i:S_i \unlhd S:\proj_i$ such that $ \bigsqcup_{i \in \omega}(\[\proj_i\];\[\inj_i\]) = \id_{\[S\]}$. 
\end{lem}
\proof We define 
$\Sigma_i = \Sigma$,
 $\nat_i = \Sigma^i \Rightarrow \Sigma$ (and so  $\inj_i = \catch_i$ and $\proj_i = \cse_i$), $(S \times T)_i = S_i \times T_i$,
and $(S \Rightarrow T)_i = S_i \Rightarrow T_i$.
\qed 
\begin{thm}For all terms $M,N$, $M \lesssim N$ if and only if $\[M\] \sqleq \[N\]$. 
\end{thm}
\proof Inequational soundness follows from soundness and adequacy: if $\[M\] \sqsubseteq \[N\]$,
 then for every context $C[\cdot]$, if $C[M]\Downarrow$ then $\[C[M]\] = \top$, by soundness,  $\[C[M]\] \sqsubseteq \[C[N]\]$ by compositionality, $\[C[N]\] = \top$, and so by adequacy $C[N]\Downarrow$ as required.

We prove inequational completeness by induction on the type of $M,N$ (closed), for which the base case is  Proposition \ref{snda}.
For example, if $M,N:S \Rightarrow T$, and $\[M\] \not = \[N\]$, then there exists $e \in \[S\]$ such that  $[M\] \spc e \not\sqsubseteq \[N\]\spc e$. Moreover, since $e = (\[\bigsqcup_{i \in \omega}(\[\lambda x. \inj_i\spc (\proj_i\spc x)\]))(e)$, by continuity there exists $i$ such that $\[M\](\[\lambda x. \inj_i\spc (\proj_i\spc x)\])(e) \not \sqsubseteq \[N\](\[\lambda x. \inj_i\spc (\proj_i\spc x)\])(e)$

By definability for $\lomt$, there is a term $L$ such that $\[L\] =
\[\proj_i\](e)$, and hence $\[M\spc (\inj_i\spc L)\] \not \sqsubseteq
\[N\spc (\inj_i\spc L)\]$. By induction hypothesis, there exists a
context $C[\_]$ such that $C[M\spc (\inj_i\spc L)\]\Downarrow$ and
$C[N\spc (\inj_i\spc L)\]\not\Downarrow$ and so $M \not \lesssim N$ as
required.\qed
 
\section{Universality for a CPS Target Language}
We have given a direct interpretation of SPCF in the category of bistable bicpos and bistable and continuous functions, but  this is in fact equivalent to a CPS (continuation-passing-style) interpretation (in the style of Streicher and Reus \cite{RS}). This may be described as a translation into a target language, $\lamw$, which is an extension of $\lomt$ with arithmetic and recursion (and which may also be used as a target calculus for CPS translation of call-by-value variants of SPCF). By proving universality for this calculus we show that it precisely captures the observably sequential functions over the given type-structure.

 Types of $\lamw$ are generated from two ground types: a data type of natural number \emph{values} and the program (or ``response'') type $\gd$. Programs of function type may take either data or programs as arguments, but must return a program --- i.e. $\nat$ may not occur on the right of an arrow. Thus the types of our language are:
$$T:: = \Na\ |\ \gd\ |\ P \times P\ |\ T \Rightarrow P$$
where $P \not = \Na$ (we refer to non-$\Na$  types as \emph{pointed}).

%We interpret $\Na$ as the ``flat'' bicpo $\coprod_{i \in \omega}\ter$.
Terms are obtained by extending the simply-typed $\lambda$-calculus (with products) with the following constants:
\begin{description} 
\item [Divergence and Error] $\bot,\e:\gd$, 
\item [Zero test]$\If:\Na \Rightarrow \Sigma \Rightarrow \Sigma \Rightarrow \Sigma$, interpreted as the function sending $0$ to $\lambda xy.x$ and $n+1$ to $\lambda xy.y$.%\footnote{From which we derive an equality test $\Eq:\Na \Rightarrow \Na \Rightarrow P \Rightarrow P \Rightarrow P = \lambda wxyz.((\If\spc (w {\mathtt =} x))\spc y)\spc z$.}
\item [Fixpoints] $\Y:(P \Rightarrow P) \Rightarrow P$, interpreted, in standard fashion, as $\bigsqcup_{i \in \omega}F^i(\bot)$, where $F = \lambda f.\lambda g.g\spc (f\spc g)$.   %\footnote{From which we derive a divergent term $\Omega:P = \Y\spc \lambda x.x$ at each pointed type.}
\end{description}
 together with 
a set of basic arithmetic constants and unary and binary operations  operations on $\Na$, including:% . % We recognize two variants of the language: in $\lamw$, we include a unary operation $\phi_f$ for every total function $f:\Na \rightarrow \Na$. We shall show that this language is sufficient to represent every bistable functional. In the effective sublanguage, $\lame$, in which we may express every computable observably sequential function, we require only the following primitive recursive  operations on $\Na$. 
\begin{enumerate}[$\bullet$]
\item zero (${\mathtt 0}$), 
 \item equality testing, $\_ = \_$, % returning zero if its arguments are equal and ${\mathtt 1}$ otherwise)
\item ``injective pairing''  $(\_*\_)$ and projections $\fst\_$ and $\snd\_$, such that $n *m >0$, $\fst(n*m) =n $ and $\snd(n*m) =t$.  
\item  a unary operation $\phi_f$ for every total function $f:\Na \rightarrow \Na$, such that $\phi_f(n) = f(n)$. 
\end{enumerate} 
\subsection{SPCF and $\lamw$}
We may embed SPCF in $\lamw$ via a fragment of the call-by-name CPS interpretation, by representing the type $\Na$  as $(\Na \Rightarrow \Sigma)\Rightarrow \Sigma$. The  constants of SPCF may thus be expressed in $\lamw$ as macros:
\begin{enumerate}[$\bullet$]
\item ${\mathtt{0}} = \lambda x.x\spc 0$
\item $\suc = \lambda f.\lambda x.f\spc \lambda n.x\spc \scc(n)$
\item $\IF = \lambda f.\lambda x.\lambda y. f\spc \lambda n.((\If\spc n)\spc x)\spc y$  
\item $\catch_n = \lambda f.\lambda x.f\spc \langle x\spc 0,\ldots x\spc (n-1)\rangle$ 
\end{enumerate}
In fact, this yields an interpretation of SPCF in the category of bistable bicpos which is equivalent to the direct one, because the objects $\Na^\top_\bot$ and $(\Na \Rightarrow \Sigma)\Rightarrow \Sigma$ are isomorphic. To show this, we extend our sequentiality result for bistable functions to those which take an argument of the form $\Na\Rightarrow D$.  Noting  that $\Na \Rightarrow D \cong \Pi_{i \in \Na}D$, if $D$ and $E$ are pointed, we say  that a function   $f:(\Na \Rightarrow D) \Rightarrow E$ is $i$-strict if $g(i) = \bot$ implies $f(g) = \bot$ and $g(i) = \top$ implies $f(g) = \top$. 
\begin{lem}\label{s1}If $A$ is pointed then every strict, continuous and bistable function $f:(\Na \Rightarrow  A)\Rightarrow \Sigma$ is $i$-strict for some $i$.
\end{lem}
\proof  
Given  $i \in \Na$, $a \in A$, and  $e \in \Na \Rightarrow A$, let $e[a]_i\in \Na \Rightarrow A$ denote the function defined:
\begin{enumerate}[$\bullet$]
\item $e[a]_i(i) = j$, 
 \item $e[a]_i(n) = e(n)$, if $n \not = i$.
\end{enumerate}
Then $\bot[\top]_i \up \bot[\top]_j$ for all $i,j$, and so by continuity and  bistability, $f(\top) = f(\bigvee\{\bot[\top]_i\ |\ i \in \Na\}) = \bigvee\{ f(\bot[\top]_i)\ |\ i \in \Na\}$, and so $f(\bot[\top]_i) = \top$ for some $i$, and $e(i) = \top$ implies $\top = f(\bot[\top]_i) \ext f(e)$. 
$\top[\bot]_i \up \bot[\top]_i$, and so $f(\top[\bot]_i) \wedge  f(\bot[\top]_i) =  f(\top[\bot]_i \wedge  \bot[\top]_i) = f(\bot) = \bot$, and so $f(\top[\bot]_i) = \bot$, and $e(i) = \bot$ implies $f(e) \ext f(\top[\bot]_i) = \bot$. 
\qed 
Hence the strict function from $\Na^\top_\bot$ to $(\Na \Rightarrow \Sigma) \Rightarrow \Sigma$ sending $\inl(n)$ to $\lambda f.f\spc n$ is an isomorphism. 
\begin{cor}$\Na^\top_\bot \cong (\Na \Rightarrow \Sigma) \Rightarrow \Sigma$. 
\end{cor} 
So every SPCF type-object is isomorphic to the corresponding $\lamw$ type-object. Moreoever, it is straightforward to show that the interpretation of SPCF constants factors through this isomorphism and hence:
\begin{prop}The direct and indirect interpretations of SPCF are equivalent.% via the isomorphism.  
\end{prop} 
\subsection{Universality for $\lamw$}
We shall now prove that every element of every $\lamw$ type-object is expressible as a term, using definable retractions.
\begin{lem}There are definable retractions from $\Na \Rightarrow \Na \Rightarrow \Sigma$ to $\Na \Rightarrow \Sigma$ and from $(\Na \Rightarrow \Sigma) \Rightarrow (\Na \Rightarrow \Sigma) \Rightarrow \Sigma$ to $(\Na \Rightarrow \Sigma) \Rightarrow \Sigma$.
\end{lem}
\proof Using the injective pairing operation, we have the embedding-projection pairs:\\ $(\lambda f.\lambda x.(f\spc \fst(x)\spc \snd(x), \lambda g.\lambda x.\lambda y.g\spc (x *y))$ and\\ $(\lambda f.\lambda x.(f\spc \lambda z.x\spc (z *0))\spc  \lambda z.x\spc (z *1), \lambda g.\lambda x. \lambda y. g\spc \lambda z.((\If\spc \snd(z))\spc (x\spc \fst(z)))\spc  (y\spc \fst(z)))$.  
\qed 
Now let $U$ be the type $\Na \Rightarrow (\Na \Rightarrow \Sigma) \Rightarrow \Sigma$. We will show that $U$ is universal amongst the (pointed) type-objects of $\lamw$ --- i.e. $T \unlhd U$ for every pointed type -- with a proof   
 based on the the sequentiality of the model. 
\begin{lem}\label{s2}If $f:U \rightarrow \Sigma$ is $i$-strict then for any $h \in U$, $f(h) = (h\spc i)\spc \lambda v.f\spc (h[\lambda y.y\spc v]_i)$.
\end{lem} 
\proof If $h(i) = \bot$, then $f(h) = \bot = (h\spc i)\spc \lambda v.(f(h[\lambda y.y\spc v]_i)$, and similarly if $h(i) = \top$. Otherwise $h(i) = \lambda p.p\spc n$ for some $n \in \Na$. Then $h[\lambda y.y\spc n]_i = h$ and so $h(i)\spc \lambda v.(f(h[\lambda y.y\spc v]_i) = f(h[\lambda y.y\spc n]_i) = f(h)$ as required. 
\qed 
Hence if $f$ is $i$-strict and  $h(i)(k) = h(i)(k')$ for all $k,k' \in \Na \Rightarrow \Sigma$, then $f(h) = h(i)(\bot)$. Note that we may express $h[a]_i$ in $\lamw$ as  $\lambda x.\If\spc x = i\spc\then\spc a\spc \el\spc (h\spc x) $.    
 % We now define a retraction from $U \Rightarrow \Sigma$ to $U$ based on this property. 
\begin{defi}Let $\inj:U \Rightarrow \Sigma = \\\Y\lambda F.\lambda f.\lambda xy.f\spc \lambda a.\lambda b.(\If\spc x\spc \then (y\spc a)\spc \el\spc ((F\spc (\lambda k.f\spc k[\lambda p.p\spc \fst(x)]_a))\spc \fst(a))\spc y$\\ and
$\proj = \Y\lambda G.\lambda g.\lambda h.(g\spc 0)\spc \lambda u.((h\spc u) \spc \lambda v.(G\spc \lambda w.g\spc (v * w))\spc h $.   
\end{defi}
To prove that this defines a retraction, we require a bound on the number of times the fixpoint must be unwound to compute $\inj(f)$ for finitary $f \in U \Rightarrow \Sigma$.
\begin{defi}A function $f: U \rightarrow \Sigma$ is $i$-dependent if there exist $g,h \in U$ such that $g(j) = h(j)$ for all $j \not = i$ and $f(g) \not = f(k)$. We shall say that $f$ has finite support if the set of $i \in \Na$ such that $f$ is $i$-dependent is finite. % has cardinality at most $n$.
\end{defi} 
\begin{lem}\label{s4}Every $f:U \Rightarrow \Sigma$ is the least upper bound of a chain of functions with finite support.
\end{lem}
\proof For each $g \in U$, define $g_i \in U$ by $g_i(x) = g(x)$ if $x \leq i$ and $g(x) = \bot$, otherwise.  Then $f^i:U \rightarrow \Sigma$, defined $f^i(g) = f(g_i)$ is continuous and bistable, and $k$-dependent only for $k \leq i$. By continuity, $\bigsqcup\{f^i\ |\ i \in \Na\} = f$.  
\qed 
\begin{lem}\label{s3}If $f$ has finitary support then $\inj(\proj(f)) = f$.  
\end{lem} 
\proof By induction on the size of the set of $n \in \Na$ such that $f$ is $n$-dependent. If $f$ is not $n$-dependent for any $n$ then it is constant, and so  $\inj(\proj(f)) = f$ by strictness of $\inj,\proj$.

Suppose $f$ is $i$-strict (hence $i$-dependent). Unfolding the fixpoint gives $\inj(f) =\\ \lambda x.\lambda y.f\spc \lambda a.\lambda b.(\If\spc x\spc \then (y\spc a)\spc \el\spc ((\inj\spc (\lambda k.f\spc k[\lambda p.p\spc \fst(x)]_a))\spc \fst(a))\spc y$.\\
Lemma \ref{s2} (on  $\lambda a.\lambda b.(\If\spc x\spc \then (y\spc a)\spc \el\spc ((\inj\spc (\lambda k.f\spc k[\lambda p.p\spc \fst(x)]_a))\spc \fst(a))\spc y$), gives 
 $\inj(f)(0)(e) =  (e\spc i)$ and $\inj(f)(m)(e) = ((\inj(\lambda k.f\spc k[\lambda y.y\spc \fst(m)]_i))\spc \fst(a))\spc e$ for $m>0$.

Hence $\proj(\inj(f))(h) \\= ((h\spc i)\spc (\lambda v.(G\spc \lambda w.\inj(f)\spc (v * w))\spc h)  \\= ((h\spc i)\spc (\lambda v.(G\spc \lambda w.\inj(\lambda k.f\spc \spc k[\fst(v * w)]_i)\spc \snd(v*w))\spc h) \\= ((h\spc i)\spc (\lambda v.(\proj\spc \lambda w.(\inj(\lambda k.f\spc \spc k[v]_i))\spc w)\spc h) \\= (h\spc i)\spc (\lambda v.(\proj\spc (\inj(\lambda k.f\spc \spc k[v]_i)))\spc h)$.

Observe that $\lambda k.f\spc \spc k[v]_i$ is $n$-dependent on strictly fewer $n$ than $f$, since it is not $i$-dependent but  if it is  $n$-dependent for some $n \not = i$ then so is $f$. Hence by hypothesis $\proj(\inj(\lambda k.f\spc k[v]_i)) =  \lambda k.f\spc k[v]_i$.  So  $\proj(\inj(f))(h) = h(i)\spc \lambda v.(f(h[v]_i) = f(h)$ by Lemma \ref{s2} as required.
\qed 

\begin{prop}$(\inj,\proj)$ form  a definable retraction from $U \Rightarrow \Sigma$ to $U$.
\end{prop}
\proof 
For each $i$, $\proj(\inj(f^i)) =f^i$ by Lemma \ref{s3}, and so $\proj(\inj(f)) = \proj(\inj(\bigsqcup_{i \in \Na}f^i)) = \bigsqcup_{i \in \Na} \proj(\inj(f^i)) =  \bigsqcup_{i \in \Na}f^i = f$.
\qed 
It is now straightforward to prove universality of $U$.
\begin{prop}For each pointed type $T$ there is a definable retraction from $T$ to $U$ $U \Rightarrow U \unlhd U$.
\end{prop}
\proof By induction on the structure of $T$. For the induction step, suppose $T = R \Rightarrow S$, then:\\
 $T \unlhd U \Rightarrow U \cong \Na \Rightarrow (\Na \Rightarrow \Sigma) \Rightarrow U \Rightarrow  \Sigma \unlhd \Na \Rightarrow (\Na \Rightarrow \Sigma) \Rightarrow U \\\cong \Na \Rightarrow \Na \Rightarrow (\Na \Rightarrow \Sigma) \Rightarrow (\Na \Rightarrow \Sigma) \Rightarrow \Sigma  \unlhd \Na \Rightarrow \Na \Rightarrow (\Na \Rightarrow \Sigma) \Rightarrow \Sigma \\ \cong (\Na \Rightarrow \Sigma) \Rightarrow \Na \Rightarrow \Na \Rightarrow \Sigma \unlhd (\Na \Rightarrow \Sigma) \Rightarrow \Na \Rightarrow \Sigma \cong U $. 
\qed 
\begin{lem}The bistable semantics of $\lamw$ is universal at type $U$.
\end{lem}
\proof Given $f\in \Na \Rightarrow (\Na \Rightarrow \Sigma) \Rightarrow \Sigma$, let  $\widehat{f}:\Na \rightarrow \Na$ be defined:
\begin{enumerate}[$\bullet$]
\item  $\widehat{f}(n) = 0$, if $f(n) = \bot$,
\item $\widehat{f}(n) = 1$, if $f(n) = \top$,
\item $\widehat{f}(n) = m+2$, if $f(n) = \inn(m)$.
\end{enumerate}
Then $f$ is definable as the term:\\
 $\lambda x.\lambda y.\If\spc \phi_{\widehat{f}}(x)\spc \then\spc \bot \spc \el\spc (\If\spc \pre(\phi_{\widehat{f}}(x))\spc \then \spc \top\spc \el\spc y\spc (\pre(\pre(\phi_{\widehat{f}}(x)))))$. 
\qed 
Hence by Lemma \ref{ret} we have shown the following.
\begin{prop}The bistable semantics of $\lamw$ is universal.
\end{prop}
%We may use the universality of $U$ to define a notion of effective computability for bistable functionals over type-objects, and show that this corresponds to definability in $\lame$. Let $\psi:(\Na \Rightarrow \Sigma) \Rightarrow \Sigma \rightharpoonup \Na$ be the partial function sending $\top$ to  $0$ and $\lambda x.x\spc n$ to $n +1$ (and undefined on $\bot$).

%\begin{prop}For any type-object $T$, an element $e \in T$ is $\lame$-definable  if and only if the partial function $\psi\cdot \inj(e):\Na \rightharpoonup \Na$ is partial recursive. 
%\end{prop}
%\proof Suppose $\psi\cdot \inj(e):\Na \rightharpoonup \Na$ is partial recursive. Then it is definable in $\lame$ as a term $M:U$ such that $M\spc n = k$ if $\psi(\inj(n)(k)) = k$, and $M\spc n = \bot$ if $\psi(\inj(n)(k))\uparrow $. Hence $e$ is definable as the term $\proj(\lambda x.\lambda y.(M\spc x)\spc \lambda z.\If\spc z\spc \then \spc \top \spc \el (y\spc \pred(z))$.

%Conversely, for any element  definable as a term $M:T$, the function  $\psi\cdot \inj(M)$ is effectively computable, by using the operational semantics of $\lame$ to evaluate $\inj(M)(n)$ for each $n$. Hence by Church's thesis, $\psi\cdot \inj(M)$ is partial recursive.
%\qed 

\subsection{Extending the Bistable Semantics}
 We may us bicpos to give (fully abstract) interpretations of  functional programming languages with a variety of features of, including recursive types, call-by-value functions, and different control primitives. In general, these models follow the same lines as those based on cpos and continuous functions. 
\begin{description}
\item[Sum Types]We may interpret sum types using either the coproduct, the ``bilifted coproduct'', $A \oplus B = (A + B)^\top_\bot$, or a ``bi-coalesced'' sum identifying the $\top$ and $\bot$ elements of its components.  Using the bilifted co-product, for example, we may  construct a fully abstract model of SPCF extended with  sums \cite{cslb}. It is straightforward to reduce full abstraction for this semantics to the case of the language without sums by using a definable retraction $A \oplus B \unlhd \nat \times A \times B$. (The injection from $A \oplus B$ to $\nat \times A \times B$  sends $\inl(e)$ to $\langle 0,e, \bot\rangle$ and $\inr(e)$ to $\langle 1,\bot,e \rangle$, and the projection from $\nat \times A \times B$ to $A \oplus B$  sends $\langle 0,d,e\rangle$ to $\inl(d)$, $\langle n+1,d,e\rangle$ to $\inr(e)$.)% and $\langle n+2,d,e\rangle$ to $\bot$.)  
\item[Recursive Types]We may interpret general recursive types using bistable variants of the standard techniques for determining colimits of $\omega$-chains of cpos  \cite{PPs,Pitts}. For example, we may  give  an observably sequential version of Plotkin's FPC \cite{FiP} by adding   recursive types to SPCF. We  may prove full abstraction for the resulting semantics by showing that every type is the limit of a chain of SPCF types, as shown for unary FPC in  \cite{ufpc}.
\item[Call-by-value]
Our constructions generalize naturally to a call-by-value setting using standard techniques; for example, the strong monad  $(\_)^\top_\bot$  meets the requirements for a model of Moggi's computational metalanguage \cite{MogC}. 

Hence we can interpret a call-by-value version of  SPCF with  $\catch$. A proof of full abstraction for this model using definable retractions is given in \cite{afflong}. Alternatively, we may interpret call-by-value SPCF with control (i.e. $\callcc$) at all types by CPS interpretation. %\emph{CPS translation} into $\lamw$.

\item[Continuation-passing style interpretation]We have given a simple interpretation of SPCF inside $\lamw$: this corresponds to a special case of the call-by-name CPS interpretation of Streicher and Reus \cite{RS}, in which (closed) terms $M:T$ are interpreted as elements of $\[T\]_c \Rightarrow \Sigma$, where $\[T\]_c$ --- the object of \emph{continuations} of type $T$ --- is defined $\[\nat\]_c = \Na \Rightarrow \Sigma$ and  $\[S \Rightarrow T\]_c = (\[S\]_c \Rightarrow \Sigma) \times \[T\]_c$. 

In general (in call-by-value, or call-by-name with sum types),    continuation-passing style interpretations will not be equivalent to those based on the lifting monad (the latter is equivalent to a \emph{linear} CPS monad \cite{LBD}). CPS interpretation  yields  models  with ``higher-order'' control ($\callcc$ at \emph{all} types), whilst lifting  yields models with ``first-order control''  ($\catch$ or $\callcc$ at ground type only). % (In the case of SPCF without sums, first-order control is sufficient to express higher-order control.)
\end{description}

\section{Bistable functions and Sequential Algorithms}
As we have already observed, sequential algorithms  also  provide a description of the fully abstract model of SPCF \cite{CCF}, and thus correspond to bistable functions. We shall now make this correspondence explicit, by showing that the set of sequential data algorithms on a sequential data structure forms a bistable bicpo, and that all bistable functions between such spaces of strategies are observably sequential, in that they are computed by a  sequential algorithm on the corresponding function-space SDS. (Similar results have been described by Curien \cite{cubi} and Streicher \cite{strd}.)

As observed in \cite{Lam,Cu}, sequential algorithms on sequential data structures  may be represented as strategies on a game (of the basic form described in \cite{AJ}). We adopt this presentation, capturing interactions which result in an error ($\top$) as  odd-length traces. %, representing divergence or ``going top'') and show that extensional order and  can be defined on the set of strategies for a game, making it a bistable biorder.

 %We  use $\varepsilon$ to refer to the empty sequence, $\sqsubseteq$  for the prefix ordering on sequences, and 
 A sequential data structure game $A$ is specified by a triple $(M_A,\lambda_A,P_A)$, where $M_A$ is a set of moves with a labelling function $\lambda_A:M_A \rightarrow \{P,O\}$ which partitions $M_A$ into sets of  Player and Opponent moves. $P_A \subseteq M_A^\circledast$ is the set of plays of $A$, where  $M_A^{\circledast}$ is the set of sequences over $A$ which are finite, alternating (i.e. P-moves are immediately preceded by O-moves and vice-versa), and contain at most one occurrence of each move and at least as many Opponent as Player moves. Key examples are the ``empty game'' $\langle \varnothing,\varnothing, \{\varepsilon\}\rangle$, and the game with one (Opponent) move $o = \langle \{o\},\langle o,O\rangle,\{\varepsilon,o\}\rangle$.

Sequential algorithms, or Player strategies on $A$, are represented as sets of plays, using odd-length sequences to represent divergences. We write $s \sqsubseteq^E t$ for the partial order on sequences defined ``$s$ is an even-length prefix of $t$ \emph{or} $s =t$''.  
\begin{defi}
 A sequential algorithm over a game $A$ is a  non-empty subset of  $P_A$, subject to the conditions:
\begin{enumerate}[$\bullet$]
 \item Even-prefix closure --- if $s \sqsubseteq^E t \in \sigma$, then $s \in \sigma$.
 \item Even-branching --- if $s,t \in \sigma$ then  $s \sqcap t \sqsubseteq^E s,t$. (So the only odd-length sequences in $\sigma$ are of maximal length.)
 \end{enumerate}
We shall write $\str(A)$ for the set of strategies over $A$. Given a strategy $\sigma$, we shall write 
$E(\sigma)$ for its set of even-length sequences (which is a strategy).
 \end{defi}
So, for instance, there are two strategies over $o$, $\{\varepsilon\}$ and $\{\varepsilon,o\}$. We shall now define an extensional order and bistable coherence making $\str(A)$ a bistable bicpo.
\begin{defi}We first  define the extensional  order on plays:\\
$s \ext t$ if $s$ is even-length and $s \sqsubseteq t$, or $t$ is odd-length and $t \sqsubseteq s$.
\end{defi} 
This is a partial order --- to show antisymmetry, note that if $s \ext t$ and $t \ext s$ then $s$ and $t$ are either both even or both odd, and hence $s = t$. Thus we may define a partial order on strategies:
$\sigma \ext \tau$ if $\forall s \in \sigma.\exists t \in \tau.s \sqleq t$. We establish that this is a partial order by proving antisymmetry.
\begin{lem}If  $\sigma \ext \tau$ and $\tau \ext \sigma$, then $\sigma = \tau$.
\end{lem}
\proof We prove that $s \in \sigma$ if and only if $s \in \tau$ by induction on length. For the induction case suppose $sab \in E(\sigma)$. Then $s \in \sigma$ and so $s \in \tau$, and there exists $t \in \tau$ such that $sab \ext t$. If  $sab \sqsubseteq t$ then $sab \in \tau$. Otherwise $t$ is odd-length and $t \sqsubseteq sab$. Then there exists $t' \in \sigma$ such that $t\ext t'$ and so $t' \sqsubseteq t\sqsubseteq sab$. But this contradicts determinacy of $\sigma$.
\qed 

So in $\str(o)$, for instance, we have $\{\varepsilon\} \ext \{\varepsilon,o\}$. More generally, for each game  there is a  $\ext$-least element $\bot$ --- the empty strategy --- and a $\sqleq$-greatest element $\top$, which contains every play consisting of at most one move.

\begin{defi}Two strategies are bistably coherent if they have the same non-divergent traces --- i.e. $\sigma \up \tau$ if $E(\sigma) = E(\tau)$. 
\end{defi}
%(So $\sigma \leq_B \tau$ if $E(\sigma) = E(\tau)$ and $O(\sigma) \subseteq O(\tau)$.) 
\begin{lem}For any game $A$, $\strat(A) = (\str(A), \ext,\up)$ is a pointed bistable bicpo. 
\end{lem}
\proof 
If $E(\sigma) = E(\tau)$ then we may define  $\sigma \wedge \tau = \sigma \cap \tau$ and $\sigma \vee \tau = \sigma \cup \tau$. These clearly satisfy the even-prefix-closure and even-branching conditions.

It is straightforward to see that $\sigma \cup \tau$ is a least upper bound, since  $\sigma,\tau \subseteq \sigma \cup \tau$, and  if $\sigma,\tau \sqleq \rho$, then for all $s \in \sigma \cup \tau$, either $s \in \sigma$ or $s \in \tau$ and we have  $r \in \rho$ such that $s \sqleq r$. 

Similarly,  $\sigma \cap \tau$ is a  lower bound ---  $\sigma \cap \tau \subseteq \sigma,\tau$. To show that it is a greatest lower bound, suppose $\rho \ext \sigma, \tau$ and $r \in \rho$. Then there exist $s\in \sigma,t\in \tau$ such that $r \sqleq s,t$. If $s$ is even-length, then $s \in E(\tau) \subseteq \sigma \cap \tau$. If $s$ is odd-length, then $s \sqsubseteq r$ and $s,t \in \sigma \cup \tau$ and so $s = t \in \sigma \cap \tau$ as required.

We now prove completeness. For a directed set of strategies $S \subseteq \strat(A)$, we define $\bigsqcup S = \{s \in P_A\ |\ \exists \sigma(s) \in S. \forall \tau \in S.\sigma(s) \ext \tau \Longrightarrow s\in \tau\}$. 

This is a well-defined strategy: if $s,t \in \bigsqcup S$ then there exists $\tau \in S$ such that $\sigma(s),\sigma(t) \ext \tau$ and so $s,t \in \tau$ and are therefore  even-branching.

$\bigsqcup S$  is an upper bound for $S$: We prove by induction on sequence length that if  $s \in \sigma \in S$ then there exists $t \in \bigsqcup S$ such that $s \ext t$. Suppose there exists $\tau \in S$ with $\sigma \ext \tau$ such that $s \not\in \tau$. Then there exists $s' \in \tau$ with $s \ext s'$, and $s'$ must be a (proper) prefix of $s$ so by hypothesis, there exists $t \in \bigsqcup S$   with $s \ext s' \ext t$.

$\bigsqcup S$ is a \emph{least} upper bound: If $\sigma \ext \tau$ for all $\sigma \in S$, then if $s \in \bigsqcup S$ then $s \in \sigma(s)\ext \tau$, and so there exists $t \in \tau$ with $s \ext t$.

$\bigsqcup$ preserves coherence: Suppose $X \up Y$, and $s \in E(\bigsqcup X)$. Then $s \in \sigma_S(s)$, and there exists $\sigma' \in X,\tau \in Y$ such that $\sigma \ext \sigma'$ and $\sigma' \up \tau$. so $s \in \sigma'$ and $s \in \tau$. If  $\tau' \in Y$ and $\tau \ext \tau'$ then either $s \in \tau'$ or else there exists odd-length $t \in \tau'$  with $t \sqsubseteq s$. But in the latter case, we may find $\sigma'',\tau''$ with $\sigma'\ext \sigma''$ and $\tau' \ext \tau''$ and $\sigma'' \up \tau''$ and so $s \in \tau''$.  Since there exists $t' \in \tau''$ with $t \ext t'$, this is a contradiction.

The proof that $\bigsqcup$ preserves bistable glbs is similar.%: Suppose $X \up Y$, and $s \in \bigsqcup X \wedge \bigsqcup Y$. Then there exists $\sigma'\in X$ and $\tau' \in Y$ such that $\sigma(s) \ext \sigma'$ and $\tau' \ext \tau(s)$ and $\sigma'\up \tau'$, so $s \in \sigma'\wedge \tau'$ 
\qed 
We shall say that a biorder arising as $\strat(A)$ for some sequential data structure is an SDS-biorder.

\subsection{Bistable Functions and observably sequential functions}

We shall now show that bistable functions between spaces of sequential algorithms correspond to sequential algorithms on the corresponding ``function-space'' sequential data structure. We follow Lamarche \cite{Lam} and Curien \cite{Cu} in decomposing this into an affine function space $A \multimap B$, and a $!$ operator.

\begin{defi}
The affine function-space $A \multimap B$ is formed as follows.
\begin{enumerate}[$\bullet$]
\item $M_{A \multimap B} = M_A + M_B$,%  $M_{A \multimap B}^- = M_A^- + M_B^+$
\item $\lambda_{A \multimap B} = [\overline{\lambda_{A}},\lambda_B]$,
\item $P_{A \multimap B} = \{t \in M_{A \multimap B}^\circledast\ |\ t \restrict A \in A\wedge t \restrict B \in B\}$. 
\end{enumerate}
We define the affine application of $\sigma:A \rightarrow B$ to  $\tau:A$:\\
$$\tau;\sigma = \{t\restrict B\ |\ t \in \sigma \wedge t \restrict A \in \tau\}$$
\end{defi}

\begin{lem}\label{af}For any sequential algorithm $\sigma:A \multimap B$, the function  from $\strat(A)$ to $\strat(B)$ sending $\tau$ to $\tau;\sigma$ is continuous and bistable.
\end{lem}
\proof
For monotonicity, suppose $\rho:A \sqleq \tau:A$. Then given $r \in \rho;\sigma$, we have $s \in \sigma$ such that $s \restrict B = r$ and $s \restrict A  \in \rho$. Hence there exists $t \in \tau$ such that $s \restrict A \sqleq t$. If $s \restrict A \sqsubseteq t$, then $s \restrict A$ is even-length and so $s \restrict A \in \tau$ and so $r = s \restrict B \in \tau$ as required.
If $s\restrict A \not \sqsubseteq t$, $t$ is odd-length and $t \sqsubseteq s\restrict A$. Hence there exists $s' \sqsubseteq^E s$ such that $s' \restrict A = t$, and $s' \restrict B \sqsubseteq s\restrict B = r$, and   $s' \restrict B$ is odd-length, so $s \restrict B \sqleq s' \restrict B$ as required.
For continuity, suppose $s \in (\bigsqcup S);\sigma$. Then there exists $t \in A \multimap B$ such that $t \restrict B = s$, and there exists  $\tau \in S$ such that $\tau \ext \tau'$ implies $t\restrict A \in \tau'$. So $\tau \ext \tau'$ implies $s \in  \tau'$ and hence $s \in \bigsqcup\{\sigma;\tau\ |\ \sigma \in S\}$.

For  bistability, we show that for all $\tau:A$, $E(\tau;\sigma) = E(E(\tau);\sigma)$.
Given $t \in  E(\tau;\sigma)$ there exists $s \in \sigma$ such that $s \restrict A \in \tau$ and $s \restrict B = t$. But since $s \restrict B$ is even-length, so are $s$ and $s \restrict A$, and therefore $s \restrict B \in E(E(\tau);\sigma)$. Preservation of bistable  lubs and glbs is straightforward. For example, if $\tau \up \rho$ then $s \in (\tau;\sigma) \cup (\rho;\sigma)$ if and only if there exists $t \in \sigma$ such that $t \restrict B = s$ and $t\restrict A \in \rho$ or $t\restrict A \in \tau$ if and only if $s \in (\rho \cup \tau);\sigma$.
\qed   
We form the game $!A$ as in \cite{Lam} by using plays of $A$ as moves of $!A$. For a sequence $s$ of such moves,  let $|s|^E = \{p \in P_A\ |\ \exists t.tp \sqsubseteq^E s\}$. 
\begin{defi}From a game $A$, we define a game $!A$ as follows:
\begin{enumerate}[$\bullet$]
\item $M_{!A} = P_A - \{\varepsilon\}$,
\item $\lambda_{!A}(sa) = \lambda_A(a)$,
\item $P_{!A} = \{s \in M_{!A}^\circledast\ |\ \forall t \sqsubseteq s. |t|^E \in \strat(A)\}$.
%\forall t \sqsubseteq^E s.|t|\cup \{\varepsilon\} \in \strat(A)\}$.
\end{enumerate}
We define the \emph{promotion} of a strategy $\sigma:A$ to a strategy $\sigma^\dag:!A$:\\ $\sigma^\dag = \{s \in P_{!A}\ |\ |s|^E \subseteq \sigma\}$. 
\end{defi}

\begin{lem}\label{dg}The function sending $\sigma$ to $\sigma^\dag$ is continuous and bistable.
\end{lem}
\proof \hfill
\begin{description}
\item[Monotonicity]We prove by induction on the length of $s$ that if  $s \in \sigma^\dag$ then there exists $t \in \tau^\dag$ such that $s \ext t$. For the induction case, suppose $s = s'(pa)$ or $s = s'(pa)(pab)$, where $s'$ is even-length. Then by hypothesis there exists $t' \in \tau^\dag$ such that $s' \ext t'$. If $t'$ is odd-length then $t' \sqsubseteq s'\sqsubseteq s$ and we are done. If $t'$ is even-length, then $s' \sqsubseteq t'$ and so $s' \in \tau^\dag$. If $s = s'(pa)$ then since $pa \in \sigma$, there must exist $q \in \tau$ with $q \ext pa$ --- i.e. $q$ is odd-length and $q\sqsubseteq pa$. Since $p \in \tau$ by even-prefix closure,  $q$ cannot be  a proper prefix of $pa$ and so $q = pa$, and so $s = s'(pa) \in \tau^\dag$. Similarly, if  $s = s'(pa)(pab)$, then either $pab \in \tau$ --- and so $s \in \tau^\dag$ --- or else $pa \in \tau$ and so $s \ext  s'(pa) \in \tau^\dag$. 
\item[Continuity]Given a directed set of strategies $S$, suppose $s \in (\bigsqcup S)^\dag$. We prove by induction on the length of $s$ that $s \in \bigsqcup\{\sigma^\dag\ |\  \sigma \in S\}$. Suppose $s = s'(pa)$. Then by hypothesis there exists $\sigma \in S$ such that $s' \in \sigma^\dag$ and $\sigma \ext \tau$ implies $s' \in tau^\dag$. Since  $|s|^E\subseteq \bigsqcup S$,  there exists $\rho \in S$ such that $\rho \ext \tau$ implies $pa \in \tau$. So there exists $\theta$ such that $\sigma,\rho \ext \theta$ and so $\theta \ext \tau$ implies $s' \in \tau^\dag$ and $pa \in \tau$ and so  $s'(pa) \in \tau^\dag$. 
\item[Bistability] Note that if $s \in P_{!A}$ is even-length then $|s|^E = E(|s|^E)$. Hence  $E(\sigma)^\dag = E(\sigma^\dag)$, and so if $E(\sigma) = E(\tau)$ then $E(\sigma^\dag) = E(\tau^\dag)$. Moreover  
$\sigma^\dag \cap \tau^\dag = \{s \in P_{!A}\ |\ |s|^E \subseteq \sigma \wedge |s|^E \subseteq \tau\} = \sigma^\dag \cap \tau^\dag$ and  $\sigma^\dag \cup \tau^\dag = \{s \in P_{!A}\ |\ |s|^E \subseteq \vee \wedge |s|^E \subseteq \tau\} = \sigma^\dag \cup \tau^\dag$.
\end{description}
\qed 
We define the application of a strategy $\sigma:A \Rightarrow B$ to a strategy $\sigma:B$ by combining the promotion and affine application operations: $\sigma \cdot \tau = \tau^\dag;\sigma$ (or directly,  $ \sigma \cdot \tau = \{s \restrict B\ |\ s \in \sigma\wedge |s\restrict \spc\spc  !A|^E \subseteq \tau\}$).  We define an observably sequential function between sequential data structures $A$ and $B$ to be a function $f:\strat(A) \rightarrow \strat(B)$ which is ``realized'' by a sequential algorithm $\sigma_f:A \Rightarrow B$ --- i.e. $f(\tau) = \sigma_f\cdot \tau$. By Lemmas \ref{af} and \ref{dg}, we have shown the following.  
\begin{prop}Every observably sequential function is continuous and bistable.
\end{prop} 
We shall now show that every strategy on $A \Rightarrow B$ corresponds to a continuous and bistable function from $\strat(A)$ to $\strat(B)$. To do so, we observe that bistable functions are \emph{stable} with respect to the inclusion order --- i.e.  continuous with respect to $\subseteq$, and conditionally multiplicative (if $\sigma,\sigma' \subseteq \tau$ then $f(\sigma \cap \sigma') = f(\sigma)\cap f(\sigma')$). % preserve intersections  of pairs of elements bounded above in $\subseteq$). 
%\begin{prop}For any SDS $A$, $(\strat(A), \ext, \subseteq)$ is a biorder in the sense of \cite{Be}.
%\end{prop}
%\proof 
%For any strategies $\sigma,\tau$, $\sigma\cap \tau$ 

%\qed 

%A function is \emph{stable} if it preserves the stable order, and is conditionally multiplicative --- 
\begin{prop}Every bistable and continuous function of SDS-biorders is stable.
\end{prop}
\proof 
Suppose $\sigma, \tau \subseteq \rho$. Let $\sigma' = (\sigma\cap \tau) \cup \{pa \in P_A \ |\ p \in E(\sigma\cap \tau)\wedge \exists q \in \sigma.pa \sqsubseteq q \wedge q \not \in \tau\}$, and $\tau' = (\sigma\cap \tau) \cup \{pa \in P_A \ |\ \exists q \in \tau.pa \sqsubseteq q  \wedge q \not \in \sigma\}$.

Then $\sigma \ext \sigma'$ (if $s \in \sigma$ then either $s \in\sigma \cap \tau \subseteq \sigma'$, or else $s'a \in\sigma'$, where $s'a$ is the maximal prefix of $s$ such that $s' \in \sigma\cap \tau$) and similarly $\tau \ext \tau'$. Moreover $\sigma'\up \tau'$ and $\sigma'\cap \tau' = \sigma \cap \tau$, and so $f(\sigma \cap \tau) = f(\sigma) \cap f(\tau) \ext  f(\sigma')\cap f(\tau') = f(\sigma'\cap \tau') = f(\sigma \cap \tau) \ext f(\sigma) \cap f(\tau)$ as required. 

Hence $f$ is also monotone with respect to $\subseteq$, and moreover continuous because every $\subseteq$-directed set is $\ext$-directed.
%Observe that $\sigma \subseteq \tau$ if and only if $\sigma \ext \tau$ and   $E(\sigma) \ext E(\tau)$: (if $\sigma \ext \tau$ and   $E(\sigma) \ext E(\tau)$ then if $s \in \sigma$, either $s \in E(\sigma)$ and so $s\in E(\tau)\subseteq \tau$ or else $s= s'a$ is odd-length, so there exists $t \in \tau$ such that $t \sqsubseteq s$. But since $s'\in \tau$, we must have $s = t$. 

%Hence if $\sigma \subseteq \tau$, $f(\sigma) \ext f(\tau)$ and $E(f(\sigma)) = $ 
\qed 
Thus each continuous and bistable function $f:\strat(A) \rightarrow \strat(B)$ has a \emph{trace}: $\tr(f)\subseteq \strat(A) \times P_B = \{(\sigma,t)\ |\  t \in f(\sigma) \wedge \forall \tau.( \sigma \uparrow \tau \wedge t \in f(\tau) \Longrightarrow \sigma \subseteq \tau)\}$. We define a sequential algorithm $\sigma_f:A \Rightarrow B$ for computing  $f$ by ``sequentializing'' this trace: $\sigma_f = \{s \in P_{A\Rightarrow B}\ | \ \forall t \sqsubseteq^E s.(|t \restrict \spc\spc !A|^E,t\restrict B) \in \tr(f)\}$.
\begin{lem}$\sigma_f$ is a well-defined strategy on $A \Rightarrow B$.
\end{lem}
\proof 
 $\sigma_f$  is even-prefix-closed by definition. To prove that it is even-branching, suppose $sab,sac \in \sigma_f$. We show that $b = c$.
\begin{enumerate}[$\bullet$]
\item If $b$ and $c$ are both  moves in $B$ then $sab\restrict B, sac \restrict C \in f(|sa \restrict \spc\spc !A|^E)$ and so $b = c$.
\item If $b$ is a  move in $!A$ and $c$ is a move in $B$ (or vice-versa), then $b$ is an odd-length sequence on $A$  and so $|sab \restrict \spc\spc !A |^E =|sa|^E\cup \{b\} \up |sa\restrict \spc\spc !A|^E = |sac\restrict \spc\spc !A|^E$. Hence $f(|sab \restrict \spc\spc !A|^E) \up f(|sac \restrict \spc\spc !A|^E)$ and so $sac \restrict B = (sa \restrict B)c \in f(|sab \restrict\spc\spc  !A|^E)$ since it is even-length. But this contradicts the assumption that the (odd-length) $sab \restrict B = sa \restrict B \in f(|sab \restrict \spc\spc !A|^E)$.
\item If $b$ and $c$ are both (Opponent) moves in $A$,  then if $b \not = c$ then $|sab \restrict \spc\spc !A |^E    \up | sac \restrict \spc\spc !A|^E$ and $|sab \restrict \spc\spc !A |^E  \wedge | sac \restrict \spc\spc !A|^E = |sa \restrict \spc\spc !A|^E$. Thus  $sa  \restrict B \in f(| sab \restrict \spc\spc !A |^E) \wedge f(| sac \restrict \spc\spc !A |^E) = f(| sab \restrict \spc\spc !A|^E) \wedge | sac \restrict\spc\spc !A |^E) = f(| sa \restrict \spc\spc !A |^E)$. But by definition of $\sigma_f$, $(|sab\restrict \spc\spc !A|^E, sab \restrict B )\in   \tr(f)$, which is a contradiction.%   This contradicts the assumption th 
\end{enumerate}
We show that if $s \in \sigma_f$ is odd-length, then there is no extension of $s$ in $\sigma_f$ by the same argument.
\qed   
We now show that the sequential algorithm $\sigma_f$ does indeed compute $f$, based on the following lemmas.
\begin{lem}Suppose $(\tau,tab) \in \tr(f)$ or $(\tau,ta) \in \tr(f)$, where $t$ is even-length, and $\sigma \subset E(\tau)$ is such that $t \in f(\sigma)$.  Then there exists a unique \emph{sequentiality index}  $\seq(\sigma)$ for $f$ at $(\sigma,t)$ --- an even-length sequence $pc \in \tau -\sigma$ such that $ta \in f(\sigma\cup\{p\})$.
\end{lem}
\proof Suppose $(\tau,tab) \in \tr(f)$ (the case $(\tau,ta)\in \tr(f)$ is similar).  In this case  $\tau = E(\tau)$  by bistability.
Let $\sigma' = \sigma \cup \{pc \in P_A\ |\ p \in \sigma \wedge  \exists d. pcd \in \tau\}$. We have $f(\sigma) \up f(\sigma')$ and so $t \in f(\sigma')$. Moreover, $\tau \ext \sigma'$, and so $f(\tau) \ext f(\sigma')$. Hence  there exists $r \in    f(\sigma')$ such that $tab \ext r$.  
If $tab \sqsubseteq r$ then $tab \in f(\sigma')$. But then $tab \in f(\sigma)$,  since $\sigma \up \sigma'$, which contradicts $\subseteq$-minimality of $\tau$. So $r \sqsubseteq tab$ is odd-length, and since $t \in f(\sigma')$, this entails $r = ta$. Since $\sigma' = \bigvee\{\sigma \cup \{pc\}\ |\ p \in \sigma\wedge \exists d.pcd \in \tau \}$, by bistability there exists a unique $pcd \in \tau$ such that $tc \in f(\sigma\cup \{pc\}) $.
\qed 
Given $(\tau,tab) \in \tr(f)$ or $(\tau,ta) \in \tr(f)$, where $t$ is even-length, we define a (finite) chain of strategies $\sigma_0 \subset \ldots \sigma_n \subseteq \tau$:
\begin{enumerate}[$\bullet$]
\item $\sigma_0 = \bigcap\{\rho \subseteq \tau\ |\ t \in f(\rho)\}$. 
\item If $\sigma_i \not = E(\tau)$,  then we define $\sigma_{i+1} = \sigma_i \cup \seq(\sigma_i)$.
\end{enumerate}
%Since the sequence is bounded and strictly increasing, there exists $n$ such that $\sigma_n = \tau$.
\begin{lem}If $\rho \subseteq E(\tau)$, $t \in f(\rho)$ and  $\seq(\rho) = pc$, then $(\rho\cup \{p\},ta) \in \tr(f)$ if and only if $\rho = \sigma_i$ for some $i$. 
\end{lem}
\proof 
%By induction on the size of $\rho$. 
Suppose $(\rho \cup\{p\},ta) \in \tr(f)$. Since there exists $n$ such that $\sigma_n = E(\tau)$, there must be some $i$ such that $\seq(\sigma_i) = pc$. Then $\seq_i \cup\{p\}$ and $\rho \cup\{p\}$ are stably coherent, and so  $tc \in f(\rho\cap \sigma_i \cup\{p\})$. But since  $(\rho \cup\{p\},ta) \in \tr(f)$, we have $\rho = \sigma_i$ as required. 

We prove the converse by induction on the size of $\rho$. Suppose $\rho = \sigma_i$, but there exists $\theta \subset \rho\cup \{p\}$ with $(\theta,ta) \in \tr(f)$. Then $t \in f(\theta - \{p\})$ (by bistability), and so by induction hypothesis $\theta -\{p\} = \sigma_j$ for some $j<i$. But then $pc$ is the sequentiality index for  $\sigma_j$, and so $pc \in \sigma_i$, which is a contradiction.
\qed 
\begin{prop}\label{p1}If $(\tau,tab) \in \tr(f)$ or $(\tau,ta) \in \tr(f)$, where $t$ is even-length, then either $(E(\tau),t) \in \tr(f)$ or else there exists (a unique) $pcd \in E(\tau)$ such that $((E(\tau) - \{pcd\}\cup\{pc\},ta) \in \tr(f)$.  
\end{prop}
\proof If $(\tau,t) \not\in \tr(f)$ then since there exists $n$ such that $\sigma_n = \tau$ $\sigma_{n+1} = \sigma_n \cup\{\seq(\sigma_i)\} = E(\tau)$, we may take $pcd =  \seq(\sigma_i)$ as required.
\qed

We may now show how to sequentialize each element of $\tr(f)$.
\begin{lem}Let  $f:\strat(A)\rightarrow \strat(B)$ be a continuous bistable function. Then for any  $(\tau,t) \in \tr(f)$, there exists a sequence $\tau \zap t \in \sigma_f$ such that $|\tau \zap t \restrict \spc\spc !A|^E = \tau$ and $\tau \zap t \restrict B = t$. 
\end{lem}
\proof By induction on the total lengths of the sequences in $\tau \cup \{t\}$.  

If $t$ is even-length and non-empty --- i.e. $t = t'ab$ --- then $\tau = E(\tau)$ by bistability and  by Proposition \ref{p1}, either $(\tau,t') \in \tr(f)$ --- and so we may define $ \tau \zap t = (\tau \zap t')ab$ ---  or there exists  $pcd \in \tau$ such that $((\tau - \{pcd\}\cup\{pc\},ta) \in \tr(f)$ --- and so we may define $ \tau \zap t =  (((\tau - \{pcd\}\cup\{pc\}) \zap ta)(pcd)b$.

Similarly, if $t$ is odd-length  i.e. $t = t'a$  --- then if $\tau = E(\tau)$, by Proposition  \ref{p1}, either $(\tau,t') \in \tr(f)$ --- and so we may define $ \tau \zap t = (\tau \zap t')ab$ ---  or there exists  $pcd \in \tau$ such that $((\tau - \{pcd\}\cup\{pc\},ta) \in \tr(f)$  --- and so we may define $ \tau \zap t =  (((\tau - \{pcd\}\cup\{pc\}) \zap ta)(pcd)$. Otherwise $\tau$ contains an odd-length sequence $q$. By minimality of $\tau$ with respect to $\subseteq$, and bistability of $f$, $q$ is unique.  By  Proposition  \ref{p1}, either $(E(\tau),t') \in \tr(f)$ --- so we may define $\tau \zap t = (E(\tau) \zap t')qa$ --- or  there exists  $pcd \in E(\tau)$ such that $((E(\tau) - \{pcd\}\cup\{pc\},ta) \in \tr(f)$  and so we may define $ \tau \zap t =  (((E(\tau) - \{pcd\}\cup\{pc\}) \zap ta)(pcd)q$.
\qed 
Thus we have shown that every bistable and continuous function $f:\strat(A) \rightarrow \strat(B)$ is observably sequential (and hence given an alternaative proof that observably sequential functions may be composed). 
%\end{cor}
\begin{prop}The SDS-biorders and observably sequential functions form a full  subcategory of $\bicp$. 
\end{prop}
%This is, moreover, a

%\proof 
%For a family of games $\{A_i\ |\ i \in I\}$ we define the product $\Pi_{i\in I}A_i$, as follows:
%\begin{enumerate}[$\bullet$]
%\item $M_{\Pi_{i\in I}A_i} = \coprod_{i \in I}M_{A_i}$, %$M_{\Pi_{i\in I}A_i}^- = \coprod_{i \in I}M_{A_i}^-$,
%\item $\lambda_{\Pi_{i\in I}A_i} = [\lambda_{A_i}\ |\ i \in I]$,
%\item $P_{\Pi_{i\in I}A_i} = \coprod_{i \in I}P_{A_i}$
%\end{enumerate} 

\section{Further Directions}
Research into bidomain models of sequential programming languages is ongoing, and includes the following themes:% Several of the themes in this article  
\begin{description}
\item[Elimination of nesting in SPCF]In \cite{aff,afflong} we use the full abstract bicpo model of SPCF to show that nested and recursive function calls in SPCF may be \emph{eliminated}: every SPCF term is observationally equivalent to one typable in an affine typing system which does not permit nesting. The proof is based on the universality of the type of first-oder functions: we show that all retractions into this type may be defined in our affine system. Since every first-order function is definable without nesting, we show that every SPCF-definable element of the model is definable in affine SPCF.   

\item [Locally Boolean Domains]We have shown that the category of sequential algorithms and sequential data structures can be fully embedded in the category of bistable bicpos and bistable and continuous functions.  This leaves open the question of how the correspondence works in the opposite direction; what is the image of the embedding, and given an object in that image, can we construct the corresponding sequential data structures? Furthermore, is there a ``linear decomposition'' of bistable bidomains into a model of linear logic, which  corresponds to that for sequential algorithms \cite{Lam,Cu}?  
In \cite{LBD} we answer these qestions by describing a notion of ``locally boolean'' domain  --- a partial order (the extensional order) with an involutive negation, which can be used to give simple definitions of the  stable and bistable orders. Our fundamental representation result for these domains is that they can all be generated (up to isomorphism) by taking products and co-products, lifting, and limits of $\omega$-chains. Hence, in particular, locally boolean domains may be viewed as games in which one player chooses indices in the product, and the other in the lifted sum.
% However, this leaves open the question of how the correspondence works in the opposite direction; what is the image of the embedding, and given an object in that image, can we construct the corresponding sequential data structures and sequential algorithms? Furthermore, is there a ``linear decomposition'' of bistable bidomains into a model of linear logic, which  corresponding to that for sequential algorithms \cite{Lam,Cu}. 
\item[Semantics of  imperative effects]Locally boolean domains form a model of linear type theory equivalent to the simple games and strategies (or affine sequential algorithms) model described by Lamarche \cite{Lam,Cu}. A more general ``linear decomposion'' of bistable functions is still under investigation. A next step is to extend our semantics beyond functional languages with control to include imperative features, non-determinism and  concurrency,
inspired by games models of functional-imperative languages such as Idealized Algol. The key to constructing such models is the identification of categorical structures shared by games and bistable models, and used to capture subtle intensional properties of such languages \cite{ctcs}. This in turn may lead to higher-order principles for reasoning about them. 

In another direction, we may obtain a semantics of  fresh  \emph{name generation} in a category of ``FM-biorders'' --- bistable biorders acted upon by the topological group of natural number automorphisms. This fits with a a natural CPS interpretation of fresh name generation given by Shinwell and Pitts to give a sequential model of a ``CPS-nu-calculus''.

\item [Other Bidomain Models] Bistable bidomains share many properties with B\'erry's original (stable) bidomains \cite{be}. This captures a different but related notion of \emph{non-deterministic} observable sequentiality, as shown by may-nand-must full abstraction results for a version of $\lamw$ with countable non-determinism \cite{fos06} (as well as fully abstract models of languages such as the lazy $\lambda$-calculus \cite{fi}). This poses the question of whether there is a general notion of bidomain embracing both stable and bistable instances, and other phenomena such as probabilistic non-determinism.

\end{description}
\section*{Acknowledgements}
Thanks to Pierre-Louis Curien and Thomas Streicher for discussions and encouragement, the referees for their comments, and Paul B. Levy for a new proof of the transitivity of $\updnarrow$.

\end{document}